\documentclass[epj]{svjour}
\usepackage{amsfonts}
\usepackage{amssymb}
\usepackage{graphics}
\usepackage[T1]{fontenc}
\usepackage{appendix}
\usepackage{enumitem}
\usepackage{booktabs}
\usepackage{multirow}
\usepackage{siunitx}
\usepackage{orcidlink}
\usepackage{arydshln}
\usepackage{caption}
\usepackage{subfig}
\usepackage{graphicx}
\usepackage{bm}

\begin{document}

\title{Late-time and Big Bang nucleosynthesis constraints for generic modify gravity surveys}

\author{N. M. Jim\'enez Cruz\inst{1,2} \and Celia Escamilla-Rivera\inst{2}
\thanks{ \email{celia.escamilla@nucleares.unam.mx}}%
}                    
\offprints{}          
\institute{Facultad de Ciencias en F\'isica y Matem\'aticas, Universidad Aut\'onoma de Chiapas. 
Ciudad Universitaria, Carretera Emiliano Zapata Km. 8, Real del Bosque (Ter\'an), 29050, Tuxtla Guti\'errez, Chiapas, M\'exico. \and Instituto de Ciencias Nucleares, Universidad Nacional Aut\'onoma de M\'exico, Circuito Exterior C.U., A.P. 70-543, M\'exico D.F. 04510, M\'exico.}
\date{Received: date / Revised version: date}
%

\abstract{
In this work, a new generic parameterisation for $f(R)$ theories is presented. 
Our proposal for a new equation of state can reproduce an $f(R)$-like evolution that describes late and early time universe within 1-$\sigma$ C.L when we use a combination of distance ladder measurements based on Cosmic Chronometers, Supernovae Ia, Baryon Acoustic Oscillation and finally, Cosmic Microwave Background and Lyman-$\alpha$ forest.
Indeed, in \cite{Jaime:2018ftn} a family of $f(R)$ cosmological viable scenarios were extensively analysed in the light of late-time measurements, were an Eos reaches a precision better than $99.2\%$ over the numerical solutions for the field equations of this theory. Moreover, in this proposal we extended the study to find constraints at the very early time that can satisfy the Big Bang Nucleosynthesis data on helium fraction, $Y_{p}$. To perform this analysis, and with our generic $w_{f(R)}$ --which can be seemed it at the same level as other parameterisations into the pipeline and analysis of observational surveys-- we consider both background and linear perturbations evolution and constrain beyond the standard $\Lambda$CDM six cosmological parameters. While there are strong constraints at background on the free parameters of our $w_{f(R)}$, we found that $f(R)$ background viable models can set early
constraints to the current Hubble constant $H_0$, which is in agreement with CMB data, but when late-time model-independent measurements are considered, $H_0$ is 
fully compatible with the $R^{H18}$ value. Finally, as an extension of these results, our proposal is capable to distinguish between $f(R)$ scenarios at both routes of the distance ladder showing a good approach to modify gravity at this level.
} 
\authorrunning{N. M. Jim\'enez and C. Escamilla-Rivera}
\titlerunning{Late-time and BBN constraints for a generic modify gravity}
\maketitle

\section{Introduction}
\label{sec:intro}

The reasons that have led to consider alternatives views of General Relativity (GR) are several and have changed over the last years. Some of them have been motivated by theoretical backgrounds, while others are predetermined by observational tests using surveys with high precision. We can thus adjoin this idea to 
the central challenge of modern cosmology, which is to shed light on the physical mechanism behind the late-time cosmic acceleration. In this line of thought, measurements have sharply improved constraints on these phenomena, and as an extension, set constraints over the free parameters of different kind of theories of gravity \cite{Clifton:2011jh}. Even though, a second challenge emerges as a consequence, the so-called $H_0$ tension, which characterises the disparity between late-time model-independent measurements and their corresponding model-dependent predictions from the early times \cite{Aghanim:2018eyx}. Statistical studies over this challenge have reinforced such issues by consider strong lensing from the H0LiCOW ($H_0$ lenses in Cosmograil's wellspring) collaboration \cite{Wong:2019kwg} and measurements from Cepheids via SH0ES (Supernovae $H_0$ for equation of state) \cite{Taubenberger:2019qna}. Furthermore,  Tip of the Red Giant Branch (TRGB Carnegie-Chicago Hubble Program) measurements have calculated a lower $H_0$ tension value \cite{Freedman:2019jwv}. 

\begin{figure*}[ht]
	\centering
	\includegraphics[width=0.7\textwidth]{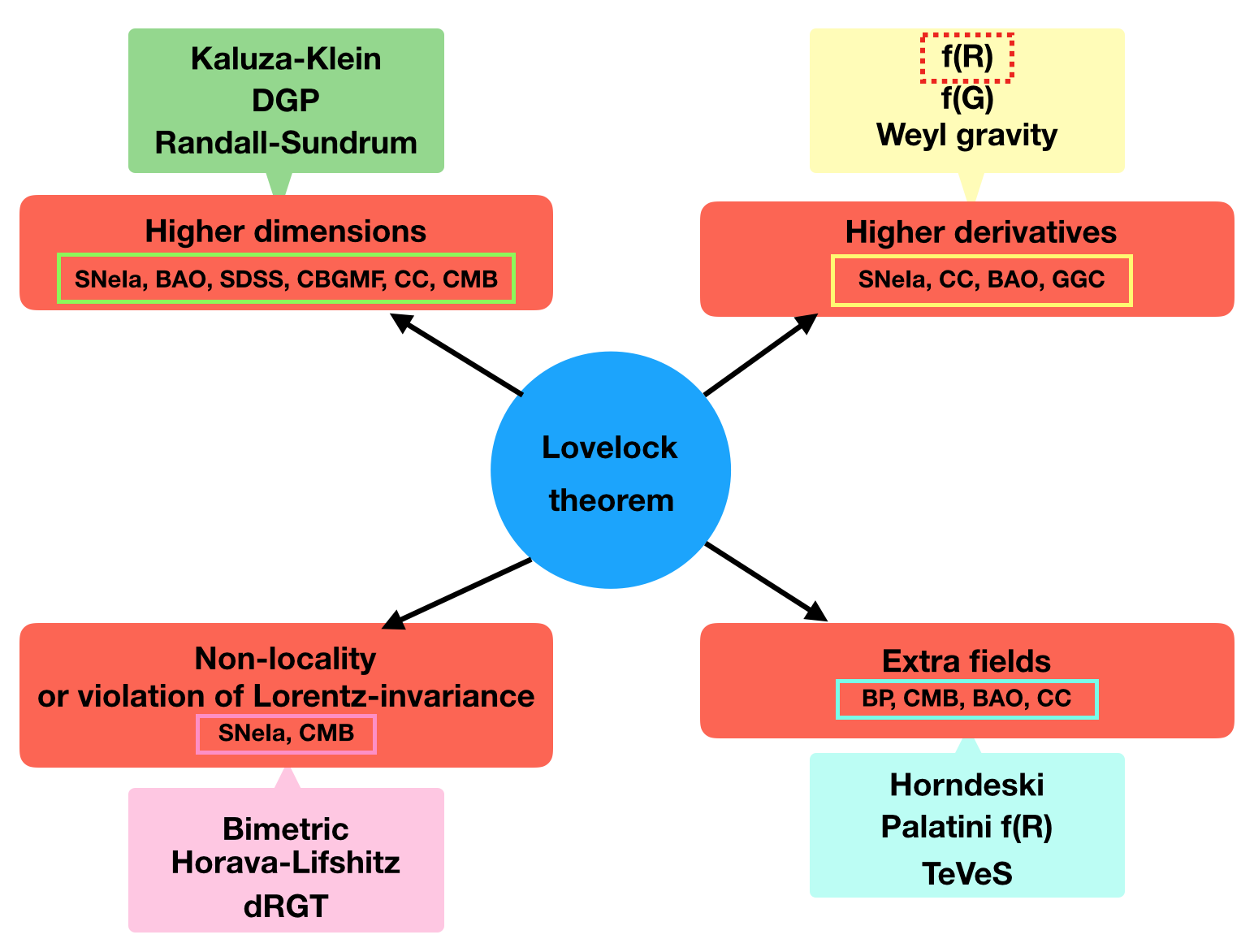}
	\caption{Architecture of theories according to violations on the Lovelock theorem. Our $w_{f(R)}$ proposal lies indicate in the dashed red color box. The color boxes indicate theories that have been constrained with observational surveys: [\textit{Green color}] SNeIa, the size of baryonic acoustic peak from Sloan Digital Sky Survey (SDSS),
	the Cluster Baryon Gas Mass Fraction (CBGMF), $H(z)$ data (CC), the radial BAO scale in the galaxy distribution and observations. [\textit{Yellow color}] SNeIa,  $H(z)$ data (CC) measurements, BAO data and several galactic globular clusters (GGC) in the Milky Way. [\textit{Pink color}] SNeIa, CMB.  [\textit{Blue light color}] Binary pulsars (BP), CMB, SDSS/BOSS and 6dF surveys  $H(z)$ data (CC)  measurements,  SNeIa, BAO SDSS.}
	\label{fig:lovelock-diagram}
\end{figure*}

To address, and even alleviate both problems, it is believed that modified gravity is an optimal path to achieve such goals. Through Lovelock's Theorem \cite{Lovelock:1972vz} we can classify theories of gravity in four different ways, depending on how we violate each of the postulates given by this theorem (see Figure \ref{fig:lovelock-diagram})
\begin{enumerate}
\item Theories that add extra fields to Einstein field equations,
\item theories that include higher order derivatives of the metric in the Action,
\item theories that add extra dimension,
\item and finally, theories with non-locality or violation of Lorentz-invariance.
\end{enumerate}

On the second classification, $f(R)$ theories of gravity \cite{Sotiriou:2008rp,Faraoni:2008mf,Capozziello:2013vna} (and references therein) are a well-behaved toy scenario to study a cosmological viable gravitational alternative. As it is standard in this kind of theory,  we can consider an arbitrary function of the Ricci scalar in the standard Einstein-Hilbert action. As an effect of introducing an arbitrary function, we can explain the late cosmic acceleration and structure formation without considering any form of dark energy or dark matter. 
Furthermore, their dynamics is kind of simple, and they can be put into correspondence with scalar-tensor theories of gravity,  from extended inflation and extended quintessence models to Kaluza-Klein and string theory. While, in the Palatini version, $f(R)$ theory also seem to have some relation with a non-perturbative quantum gravity. 

Coming back to the tension issue, some modified gravity theories (e.g. Galileon \cite{Nicolis:2008in,Escamilla-Rivera:2015ova}, Dvali, Gabadadze, and Porrati (DGP) cosmologies \cite{Dvali:2000hr,Barcenas-Enriquez:2018ili} just to cite a few) may reconcile the Planck with high $H_0$ values \cite{Barreira:2014jha}, although it is known that such models have problems with either cosmology \cite{Renk:2017rzu} or gravitational waves \cite{Ezquiaga:2017ekz}. 
Our first goal is to focus our attention in $f(R)$ gravity models, which do not suffer from the above issues and
their characteristic EoS \cite{Jaime:2013zwa} makes them an appealing framework in order to reproduce the dynamical evolution of an EoS $\omega(z)$ found in \cite{Zhao:2017cud}. In this context, the resulting field equations are of fourth order on the metric and behave like attractors; therefore their implementation in surveys, or in N-body, and Boltzmann codes requires many assumptions. In this case, adapting an entire dynamical system of fourth-order equations that can describe a viable $f(R)$ theory is not an easy task. Moreover, if we found a dynamical EoS that can reproduce a variety of $f(R)$ models between $[0.5\%-0.8\%]$ of precision can help to study these models in a straightforward way. This proposal was developed in \cite{Jaime:2018ftn} with the so-called JJE parameterisation and tested with late-time surveys. As a first goal, reasonable results were obtained since they alleviate the tension between observations coming from different stages of the universe, and of course, the adaptation of a \textit{generic} EoS coming from a numerical result of a viable $f(R)$ theory is easer and with less degeneracy in the specific form of $f(R)$. 

Our second goal is to study these alternative $f(R)$ EoS approach to choose a consistent cosmological model capable to reproduce a viable
cosmic history. This idea is directly connected to extremas of the distance ladder: the early (e.g inflation scenarios) and late-time universe (e.g. dark energy era) with their corresponding large scale structure formation era. On one hand, in the late-time route of the distance ladder with SNeIa we can anchor their relative distances with an absolute distance measurement \cite{DiValentino:2020vhf,DiValentino:2020zio}, taking into account that supernova distances on their own are not absolute distances because of the unknown intrinsic luminosity. Nonetheless, these objects can map the late cosmic expansion history of the universe and with anchored distances, the determination of cosmological models with insensitive absolute distance can be possible. The latter can be done by considering distances measured from baryon acoustic oscillations (BAO) and galaxy ages cosmic chronometers (CC).

For higher dimensions theories have been tested using data sets from SNeIa, the size of baryonic acoustic peak from Sloan Digital Sky Survey (SDSS) compilations, 
the Cluster Baryon Gas Mass Fraction (CBGMF) \cite{Movahed:2007ie}, $H(z)$ data, the radial BAO scale in the galaxy distribution \cite{Barcenas-Enriquez:2018ili}, and observations obtained by the Planck Collaboration (\cite{Milosevic:2018gck}, \cite{Loc:2020mrn}). For higher derivative theories, analyses have been performed using SNeIa and Cosmic Chronometers measurements \cite{DAgostino:2019hvh}, BAO data (\cite{Dev:2008rx}, \cite{Nunes:2016drj}) and several galactic globular clusters (GGC) in the Milky Way \cite{Islam:2018ymd}. For the non-locality or violation of Lorentz invariance theories again have been employed SNeIa \cite{Luben:2020xll}, CMB (\cite{vonStrauss:2011mq}, \cite{Nilsson:2018knn}, \cite{Dutta:2009jn}, \cite{Amendola:2019fhc}). Finally, in the case of extra fields theories tests have been done with binary pulsars (BP) \cite{Avdeev:2018ihq},  CMB, SDSS/BOSS and 6dF surveys \cite{Noller:2018eht}, $H(z)$ measurements,  SNeIa as well as BAO peak in the SDSS luminous red galaxy sample and the CMB shift parameter (\cite{Fay:2007gg}, \cite{Pires:2010fv}). All these observations have set constraints on the parameters of each of the models described in Figure \ref{fig:lovelock-diagram}.

On the other hand, in the early-time route of the distance ladder, Big Bang Nucleosynthesis (BBN) allows a technique to measure precise abundances of hydrogen, helium, lithium and deuterium to test cosmological model being extremely sensitive at early times. In this scenario, cosmological models can be tested to obtain a significant $\Omega_b$ and the mass fraction of $^{4}$He ($Y_P$) inferred from the BBN combined with the baryon density from the Cosmic Microwave Background (CMB) \cite{Aver:2015iza}. 

With the above ideas, in this work we will propose a generic $w_{f(R)}$ that can be modelling to obtain the current observed cosmic accelerated expansion. With this EoS at hand, a perturbative analysis can be performed to test $f(R)$ viable scenarios from the late-time universe (i.e. using SNeIa, BAO, CC data) to early-time universe (i.e using BNN, CMB and Lyman-$\alpha$)

This paper is organised as follows: 
in Sec. \ref{sec:theories} we summarise the motivation for the geometric equation of state from $f(R)$ theories 
and describe those cosmologically viable models.
In Sec. \ref{sec:generic_JJE} we introduce our proposed $w_{f(R)}$ EoS parameterisation. Also we will detail how its cosmological
evolution has a  precision better than  $99.2\%$ and $99.5\%$. 
In Sec. \ref{sec:JJE_pertb} we develop the linear perturbation analysis for our proposed  $w_{f(R)}$ EoS parameterisation.
In Sec. \ref{sec:surveys} we describe the surveys used along this work. 
In Sec. \ref{sec:method-late} and \ref{sec:method-early} we present the results for the late-time and early time analysis of the
data fitting process, respectively. 
Finally, in Sec. \ref{sec:conclusions} we discuss our results from the full sample point of view to present a generic combo of cosmological constraints for $f(R)$ viable models.


\section{Viable cosmologies in $f(R)$ theories}
\label{sec:theories}

To modify theories of gravity, as we detailed in Fig. \ref{fig:lovelock-diagram}, we add to the Einstein-Hilbert action an arbitrary function $f(R)$
of the Ricci scalar $R$ following:
\begin{equation}
\label{eq:f(R)}
S[g_{\mu\nu},{\mbox{\boldmath{$\psi$}}}] =
\!\! \int \!\! \frac{f(R)}{2\kappa} \sqrt{-g} \: d^4 x 
+ S_{\rm m}[g_{\mu\nu}, {\mbox{\boldmath{$\psi$}}}] \; ,
\end{equation}
where $G=c=1$ and $\kappa \equiv 8\pi$. $S_{\rm m}[g_{\mu\nu}, {\mbox{\boldmath{$\psi$}}}]$ is the 
action for matter. $f(R)$ denotes a function of the Ricci scalar $R$. Computing the variation of (\ref{eq:f(R)}) we can obtain the field equations
\begin{eqnarray}
\label{fieldeq3}
& G_{\mu\nu} =& \frac{1}{f_R}\Bigl{[} f_{RR} \nabla_\mu \nabla_\nu R +
 f_{RRR} (\nabla_\mu R)(\nabla_\nu R) 
-\frac{g_{\mu\nu}}{6}\Big{(} Rf_R+ f + 2\kappa T \Big{)} 
+ \kappa T_{\mu\nu} \Bigl{]}, \; 
\end{eqnarray}
where $G_{\mu\nu}= R_{\mu\nu}-g_{\mu\nu}R/2$ is the Einstein tensor. To obtain $f(R)$ models that can be cosmologically viable (i.e. models which predict a matter dominated period followed by an accelerated expansion) we need a flat, homogeneous and isotropic space-time given by a Friedman-Lema\^itre-Robertson-Walker (FLRW) metric
$ds^2 = - dt^2  + a^2(t)\!\left[ dr^2 + r^2 \left(d\theta^2 + \sin^2\theta d\varphi^2\right)\right]$.
Under these assumptions, we can consider an energy momentum tensor for a fluid composed by baryons, dark matter and radiation. With all these, we can derive a second order differential equation for the Ricci scalar

\begin{eqnarray}
\label{traceRt}
& \ddot R = &-3H \dot R -  \frac{1}{3 f_{RR}}\left[ 3f_{RRR} \dot R^2 + 2f- f_R R + \kappa T \right], \,\,\,\,\,\, \\
\label{Hgen}
& H^2 = & -\frac{1}{f_{RR}}\left[f_{RR}H\dot{R}-\frac{1}{6}(Rf_{R}-f) \right]-\frac{\kappa T^{t}_{t}}{3f_{R}}, \\
\label{Hdotgen}
& \dot{H}= & -H^2 -\frac{1}{f_{R}} \left[ f_{RR}H\dot{R} + \frac{f}{6}+\frac{\kappa T^{t}_{t}}{3} \right]  \,\,\,,
\end{eqnarray}
where $H = \dot a/a$. Under this consideration, we derive an optimal form of the density and pressure for this effective fluid \cite{Jaime:2012gc}:
\begin{eqnarray}\label{eq:density-pressure}
\rho_{\text{eff}} &=&\frac{1}{\kappa f_R}\left[\frac{1}{2}(f_R R-f) - 3f_{RR} H \dot{R} +\kappa \rho (1-f_R)\right], \label{eq:density}\\
P_{\text{eff}}&=& -\frac{1}{3\kappa f_R} \left[\frac{1}{2}(f_R R +f) +3f_{RR} H\dot{R} -\kappa (\rho -3p f_{R})\right]. \label{eq:pressure}
\end{eqnarray}
where the EoS for this kind of fluid in $f(R)$ is \footnote{Here we consider a Ricci scalar approach to $f(R)$ \cite{Jaime:2010kn} and then a $P$ and $\rho$, of the matter and radiation content
according to \cite{Berti:2015itd}}
\begin{equation}
\label{eq:GEoS}
w_{\text{eff}} = \frac{3H^2-3\kappa P-R}{3(3H^2-\kappa\rho)}.
\end{equation}

At this point, among a extensive forms of modify gravity models, we can denote that \textit{a $f(R)$ model is cosmologically acceptable when they have a standard matter stage followed by an
accelerated attractor}. In this line of thought, the most successful $f(R)$ models that fulfill these main characteristics are 
\begin{enumerate}
 \item Starobinsky model \cite{Starobinsky:2007hu}
          	 \begin{equation}\label{eq:f(R)-St}
          	 f(R)= R+\lambda R_{S}\left[ \left( 1+\frac{R^2}{R^2_{S}}\right)^{-q}-1\right],
			 \end{equation}     
			 
 \item Hu-Sawicki model  \cite{Hu:2007nk} ~\footnote{Here the parameters $c_1$ and $c_2$ are related with $f_R^0$ and $\Omega_m^0$ according to \cite{Hu:2007nk}.} 
   			 \begin{equation}\label{eq:f(R)-HS}
   				 f(R)= R- R_{\rm HS}\frac{c_{1}\left(\frac{R}{R_{\rm HS}}\right)^n}{c_{2}\left(\frac{R}{R_{\rm HS}}\right)^n+1} \; ,
    		 \end{equation}
   		
		 \item The Exponential model \cite{EXP} 
             \begin{equation}\label{eq:f(R)-exp}
             f(R)=R+\beta R_* (1-e^{-R/R_*}).
             \end{equation}                  	  
\end{enumerate}
All the free parameters involved in these models should be constrained according to observations, this can be achieved by integrating the field equations (\ref{traceRt})-(\ref{Hgen})-(\ref{Hdotgen}) from the past to the future. In a certain approximation, the above models provide an accelerated behaviour with a $w_{\text{eff}} \approx -1$, analogous to $\Lambda$CDM. Moreover, 
in (\ref{eq:f(R)-HS}) and (\ref{eq:f(R)-St}) such behaviour goes asymptotically to a deSitter critical point $R_{(z \rightarrow -1)} > 0$. For (\ref{eq:f(R)-exp}) the future evolution is asymptotically $R_{(z \rightarrow -1)} = 0$, with a long enough accelerated stage.


\section{A generic equation of state for modify gravity}
\label{sec:generic_JJE}

As it was mentioned, the possibility to explain the late-time accelerated cosmic expansion in $f(R)$ theories, usually an effective fluid is under consideration. The EoS that describes this kind of fluid is a standard expression that mimics dark energy in the form of parameterisations that simply show deviations from the $\Lambda$CDM model. The degeneracy of proposals along this idea has lead us to motivated the study of alternative models of the universe from a \textit{generic} EoS point of view that can reproduce a viable cosmology as we described. 
Hereby, in \cite{Jaime:2018ftn} was proposed a new parameterisation that can provide a useful way to implement a $f(R)$-like cosmology in observational precision tests, surveys or even in Boltzmann codes. The so-called JJE parameterisation is based on the numerical results coming from the integration of the field equations in (\ref{traceRt})-(\ref{Hgen})-(\ref{Hdotgen}) and given by
\begin{equation}
\label{eq:fit}
w(z) _{\text{JJE}}= -1 + \frac{w_0}{1+w_1z^{w_2}}\cos(w_3+z).
\end{equation}
where $\omega_i$ (i=1,2,3) runs the free parameters and $z=a_0/a-1$. Eq. (\ref{eq:fit}) has a current value $w(z=0) = w_0\cos(w_3)-1$, can recover $w_{\text{eff}}=-1$ at large $z$, and allows dynamics (oscillations) in the range of observations. From this proposal we can obtain our three $f(R)$ models: (\ref{eq:f(R)-HS})-(\ref{eq:f(R)-St}) within a $0.5\%$, and (\ref{eq:f(R)-exp}) within a $0.8\%$ of precision, which are reasonable values where current astrophysical surveys can fix a cut-off over the cosmological constraints.

Moreover, finding the best fit values of the free parameters of (\ref{eq:fit}), it is a difficult task and opens the possibility to have degeneracy at this level. Therefore, in this work we implement this parameterisation in a modified version of CLASS code to avoid degeneracy in the periodicity over the cosine function and rewriting a new parameterisation by performing a shift in the argument of this function as 
\begin{equation}
\label{eq:JJEeos}
w_{f(R)}=-1+\frac{\omega_{0}\cos{(\alpha \nu (z))}}{1+ \omega_{1}  z^{\omega_{2}} },
\end{equation}
where
$\nu (z)=\frac{2 \pi}{(\sqrt{6}z+1)^{1/2}}$. Additionally, we fixed the values for the free parameters of the models according if they pass or not e.g Solar System tests, 
otherwise, the integral $3\int (1+w_{JJE})da/a$, can end up calculating no analytical or even no numerical (convergent) solutions.

\begin{figure*}[ht]
	\centering
		\includegraphics[width=0.47\textwidth]{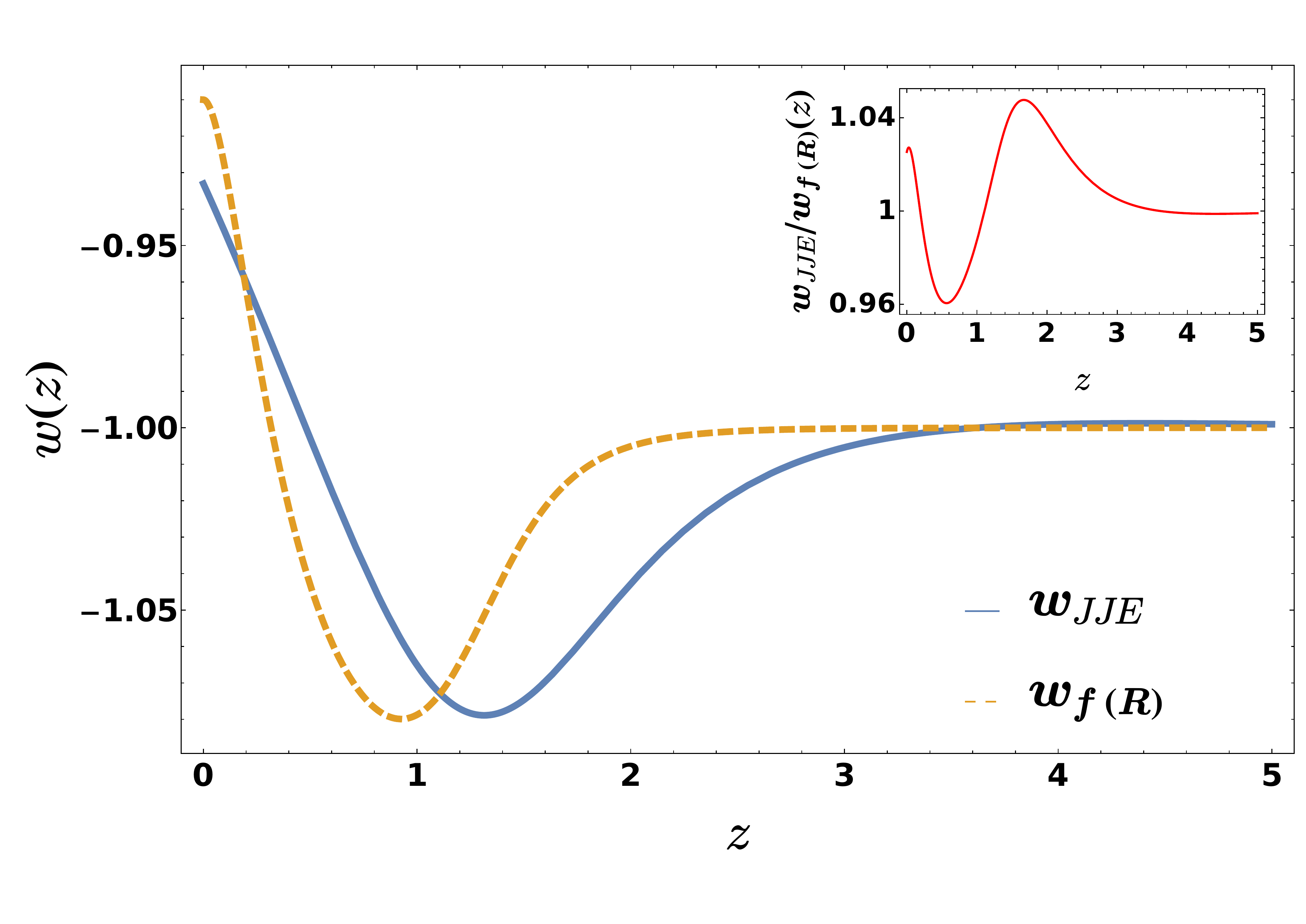}
		\includegraphics[width=0.47\textwidth]{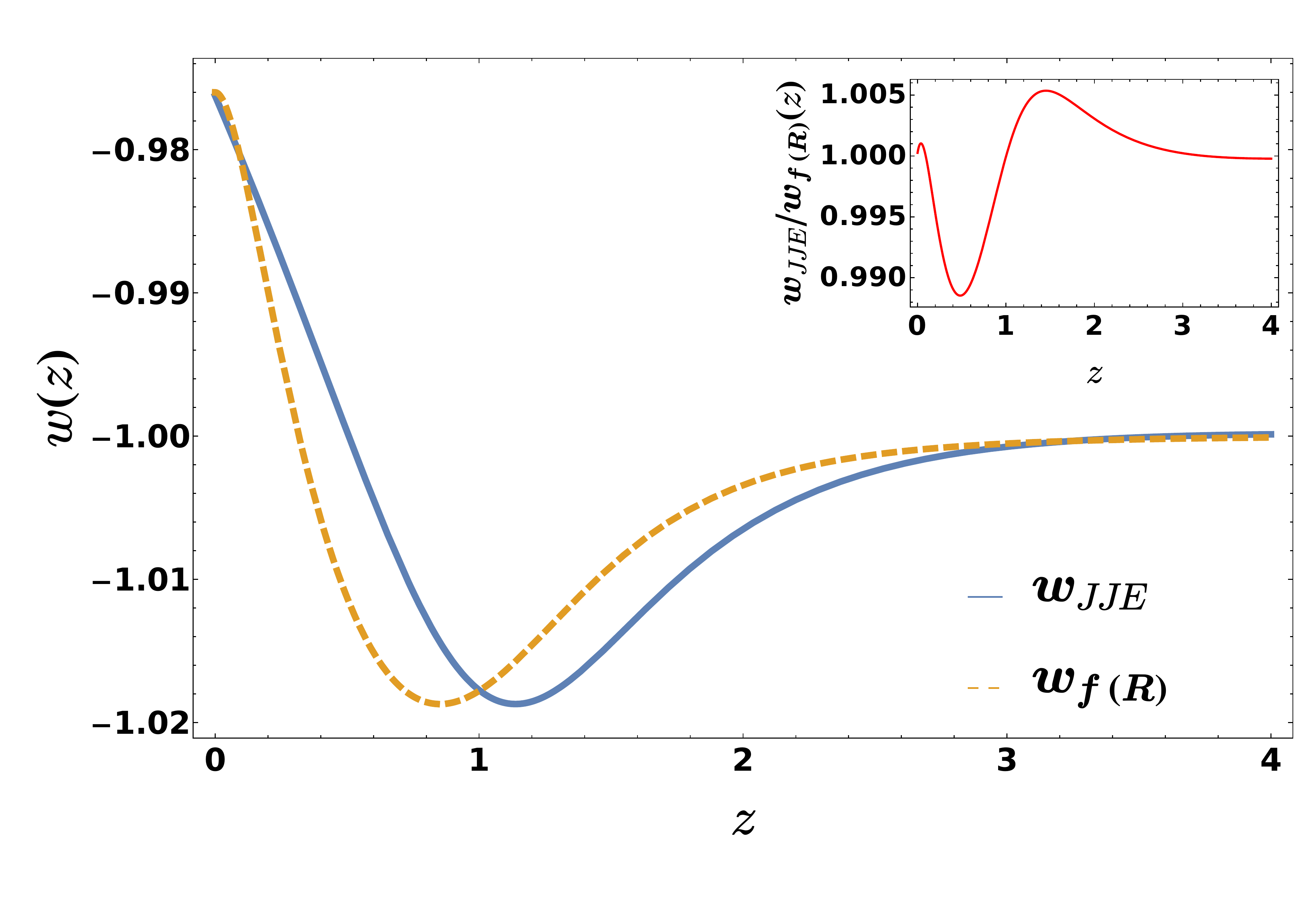}
		\includegraphics[width=0.47\textwidth]{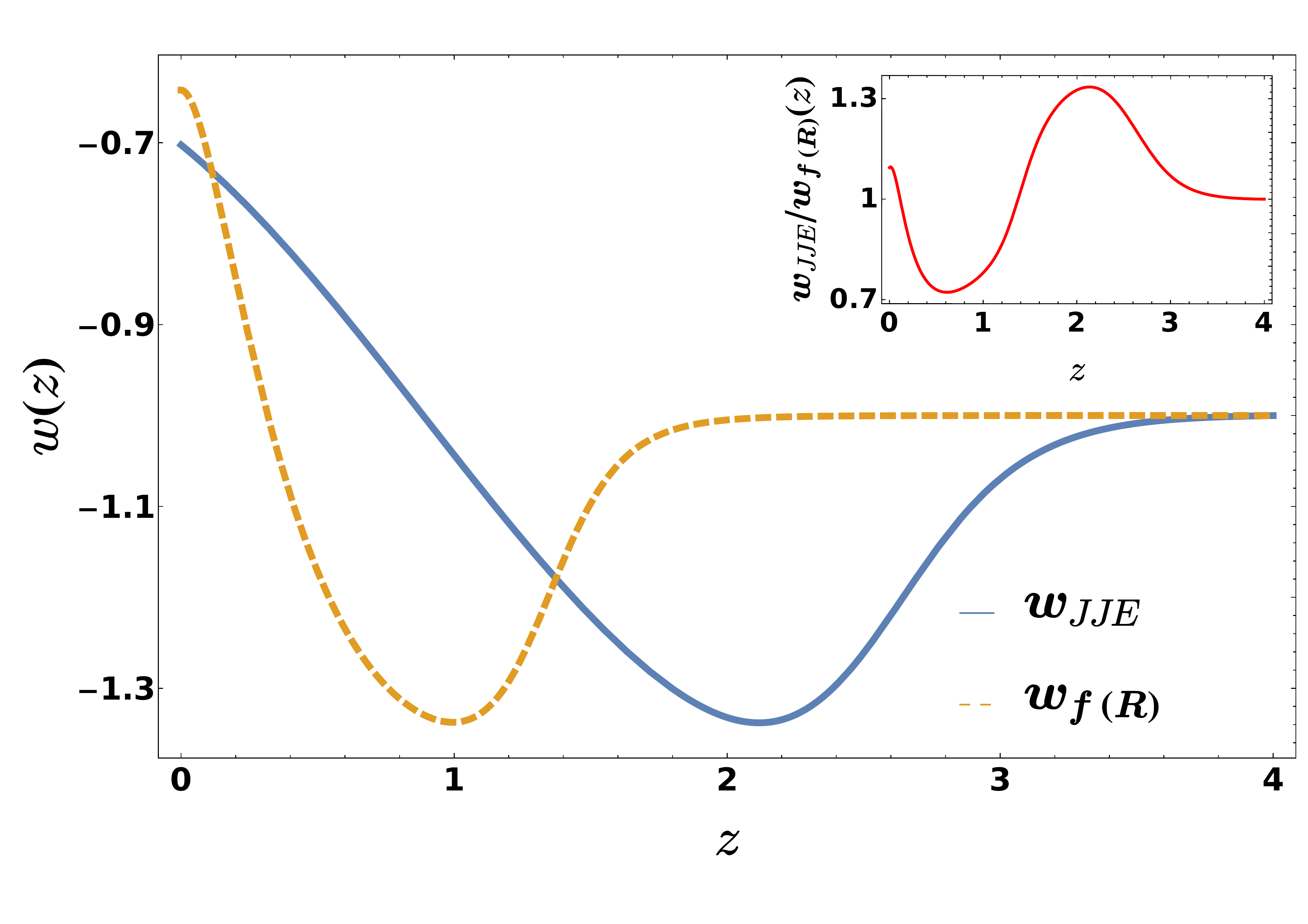}
	\caption{Evolution of Eq. (\ref{eq:fit}) (blue solid line) and Eq.(\ref{eq:JJEeos}) (orange dashed line) according to the bestfits detailed in Table \ref{tab:fit_num}. The inner plots denote the comparison between EoS as $w_{\text{JJE}} / w_{f(R)}$. 
	\textit{Top Right:} Starobinsky model. For this model notice that there is a variation of 4\% between $w_{\text{JJE}} $ y $w_{f(R)}$. 
		\textit{Top Left:} Hu-Sawicki model. For this model notice that there is a variation of [0.5\% - 1\%]  between $w_{\text{JJE}} $ y $w_{f(R)}$. 
				\textit{Bottom Middle:} Exponential model. For this model notice that there is a variation of  30\%  between $w_{\text{JJE}} $ y $w_{f(R)}$. }
	\label{fig:approx_models}
\end{figure*}

After introducing (\ref{eq:JJEeos}) parameterisation to CLASS code, we perform a comparison between models described. The best fit values are detailed in Table \ref{tab:fit_num} and showed in Figure \ref{fig:approx_models}. 

\begin{table*}
\begin{center}
\caption{Best fits obtained from the fit between $w_{JJE}$ (\ref{eq:fit}) and $w_{f(R)}$ (\ref{eq:JJEeos}). We consider as an ansatz a flat prior of $\Omega_{m}=0.3$.}
\begin{tabular}{|c|c|c|c|c|c|c|}
 \hline
 & \multicolumn{2}{ |c| }{\textbf{Starobinsky model}} & \multicolumn{2}{ |c| }{\textbf{Hu - Sawicki model}} & \multicolumn{2}{ |c| }{\textbf{Exponential model}} \\ \hline
Parameter & $w_{JJE}$ & $w_{f(R)}$ & $w_{JJE}$ & $w_{f(R)}$ & $w_{JJE}$ & $w_{f(R)}$ \\ \hline
$\omega_{0}$ & $0.145$ & $0.09$ & $0.049$ & $0.024$ & $0.384$ & $0.358$ \\
$\omega_{1}$ & $0.106$ & $0.11$ & $0.310$ & $0.310$ & $0.000014$ & $0.03$ \\
$\omega_{2}$ & $4.491$ & $7$ & $4.395$ & $4.000$ & $11.370$ & $11.00$  \\
$\omega_{3}$ & $7.374$ & $-$ & $7.348$ & $-$ & $0.684$ & $-$  \\
$\alpha$ &$-$ & $1$ &$-$ & $1$ &$-$ & $1$  \\
\hline
 \end{tabular} \label{tab:fit_num}
 \end{center}
 \end{table*}

\section{Perturbations with the generic $w_{f(R)}$}
\label{sec:JJE_pertb}

Cosmological information comes from several sources. However, at present time, the CMB dominates the constraints on the standard model of cosmology. This era of precision cosmology can be seen through plots of the current status of the power spectra coming from the CMB. Furthermore, this spectrum carries relevant information on both the early and late time universe. 
We took the advantage our modified version of CLASS code\footnote{Another example of numerical code is hiCLASS, which can compute predictions for modified models. However, our new code can consider any \textit{additional} modification as an effective fluid derived from the numerical solution of the field equation in a generic way.} to analyze the perturbations of our proposed $w_{\text{f(R)}}$ Eq. (\ref{eq:JJEeos}), which as we see in Figure \ref{fig:approx_models} can reproduce three viable $f(R)$ models. This implementation allows to analyze the structure formation since each spectra (power spectrum) is related to the curvature, baryonic matter, dark matter, etc. 

In this work we started with the perturbations obtained for $\Lambda$CDM and afterwards we carry out a modified version that includes our proposal  (\ref{eq:JJEeos}) as an effective-like fluid. The effects will be seen as deviations in the power spectrum. These perturbations are described by equations for density contrast and velocity divergence in the synchronous gauge and valid for a perfect fluid \cite{Ma:1995ey,Hu:1998kj}
 \begin{eqnarray}
 \dot{\delta}_{i} + 3\mathcal{H}(c^2_{s,eff} - \omega_{i})\left[\delta_{i}+3\mathcal{H}(1+\omega_{i})\frac{v_{i}}{k}\right]+(1+\omega_{i})k v_{i}+ 3\mathcal{H} \dot{w}_i v_i/k &=& -3(1+w_i)\dot{h}  ,\label{eq:idelta_cmb}\\
\dot{v}_i + \mathcal{H}(1-3c^2_{s,eff})v_{i} &=& \frac{k \delta_i c^2_{s,eff}}{1+\omega_{i}}\label{eq:iv_cmb}.
 \end{eqnarray}
The derivatives denoted by dots are with respect the conformal time, $\mathcal{H}$ is the conformal Hubble parameter, $\omega_i = w_{f(R)}$,  the $c^2_{s,eff}$ is the effective sound speed in the rest frame and $v_i$ is the velocity of the $i$th fluid. To avoid the crossing instability problem, we use the Parameterised Post-Friedmann (PPF) approach in CAMB code \footnote{The Boltzmann solver code to compute the evolution of linear perturbations available at \url{https://camb.info}}. 

\section{Observational  samples}
\label{sec:surveys}

Given that we are interested in modelling the early and late-time cosmic evolution we use current observational data from SNeIa luminous distance, BAO redshift surveys, the latest high-z measurements of $H(z)$ from Cosmic Chronometers (CC), CMB data from Planck 2018 and measurements via the detection of Lyman-$\alpha$ (Ly$\alpha$) radiation in both emission and absorption lines. We described each of them below.

\begin{itemize}

\item[(a)] Cosmic Chronometers.
Consist in passively evolving old galaxies whose redshifts are known. For this sample, the expansion history of the universe can be computed directly from their differential ages. According to this, we consider the current data described in Table \ref{tab:cc}.

\begin{table*}[h]
\centering
\begin{tabular}{lllllllll}
\hline
$z$    & $H(z)$ & $\sigma_{H(z)}$ & $\quad$ & $z$   & $H(z)$ & $\sigma_{H(z)}$  \\
\hline
0.07   & 69.0 & 19.6        & & 0.4783 & 80.9   & 9.0   \\
0.09   & 69.0 & 12.0        & & 0.48   & 97.0   & 62.0  \\
0.12   & 68.6 & 26.2       & & 0.593  & 104.0  & 13.0 \\
0.17   & 83.0 & 8.0       & & 0.68   & 92.0   & 8.0      \\
0.179  & 75.0 & 4.0   & & 0.781  & 105.0  & 12.0  \\
0.199  & 75.0 & 5.0      & & 0.875  & 125.0  & 17.0  \\
0.2    & 72.9 & 29.6         & & 0.88   & 90.0   & 40.0\\
0.27   & 77.0 & 14.0      & & 0.9    & 117.0  & 23.0  \\
0.28   & 88.8 & 36.6      & & 1.037  & 154.0  & 20.0  \\
0.352  & 83.0 & 14.0     & & 1.3    & 168.0  & 17.0 \\
0.3802 & 83.0 & 13.5     & & 1.363  & 160.0  & 33.6  \\
0.4    & 95.0 & 17.0     & & 1.43   & 177.0  & 18.0  \\
0.4004 & 77.0 & 10.2    & & 1.53   & 140.0  & 14.0  \\
0.4247 & 87.1 & 11.2     & & 1.75   & 202.0  & 40.0  \\
0.4497 & 92.8 & 12.9   & & 1.965  & 186.5  & 50.4  \\
0.47   & 89.0 & 49.6&  & &  \\
\hline
\end{tabular}
\caption{Cosmic Chronometers data sample from  \cite{Marra:2017pst}.}
\label{tab:cc}
\end{table*}

\item[(b)] Pantheon Supernovae Type Ia.
The current sample is Pantheon compilation in 40 bins\footnote{\url{https://github.com/dscolnic/Pantheon}}.
The standard description provide values of the distance modulus $\mu$, which can be directly employed to compute the luminosity distance $d_{L}$ (in Mpc) according to
$\mu\left(z\right)=5\log\left[\frac{d_{L}\left(z\right)}{1 \text{ Mpc}}\right]+25\,.$

We add to this quantity the nuisance parameter $M$, which is related (and degenerate) to the prior on $H_{0}$. In this work we consider additional priors to calibrate this sample. Assuming spatial flatness, $d_{L}$ is related to the comoving distance $\mathcal{D}$ as
$\mathcal{D}\left(z\right)=\frac{H_{0}}{c}\left(1+z\right)^{-1}10^{\frac{\mu\left(z\right)}{5}-5}\,.$
This expression can be normalised by the Hubble function $E\left(z\right)\equiv H\left(z\right)/H_{0}$ to obtain
$\mathcal{D}\left(z\right)=\int_{0}^{z}\frac{d\tilde{z}}{E\left(\tilde{z}\right)}$.
And we compute the best fits by minimizing the quantity
\begin{eqnarray}
\chi_{\text{SN}_{\text{Pantheon}}}^2
=\sum^{N_{\text{Pantheon}}}_{i=1}{\frac{\left[\mu(z_i ,\Omega_m ;\mu_0,
w_i)-\mu_{\text{obs}}(z_i)\right]^2}{\sigma^{2}_{\mu,i}}}
\end{eqnarray}
where $N=1048$ is the number of SN observations, $w_i$ denotes the free parameters in (\ref{eq:JJEeos}) and $\sigma^{2}_{\mu,i}$ are the measurements variances.

\item[(c)] BAO sample.
Also we consider model-independent angular  BAO 
determinations from the angular correlation function. A total of 
14 uncorrelated data points are reported in Table~\ref{tab-bao}.
\begin{table*}[t]
\centering
\begin{tabular}{lllll}
\hline
Catalog & $z$    & $\theta(z)$ & $\sigma_{\theta(z)}$ \\
\hline
SDSS-DR7          & 0.235        & 9.06        & 0.23        \\
SDSS-DR7          & 0.365        & 6.33        & 0.22      \\
SDSS-DR10         & 0.450        & 4.77        & 0.17     \\
SDSS-DR10         & 0.470        & 5.02        & 0.25      \\
SDSS-DR10         & 0.490        & 4.99        & 0.21       \\
SDSS-DR10         & 0.510        & 4.81        & 0.17     \\
SDSS-DR10         & 0.530        & 4.29        & 0.30    \\
SDSS-DR10         & 0.550        & 4.25        & 0.25      \\
SDSS-DR11         & 0.570        & 4.59        & 0.36      \\
SDSS-DR11         & 0.590        & 4.39        & 0.33       \\
SDSS-DR11         & 0.610        & 3.85        & 0.31        \\
SDSS-DR11         & 0.630        & 3.90        & 0.43        \\
SDSS-DR11         & 0.650        & 3.55        & 0.16       \\
SDSS-DR12Q$\quad$ & 2.225$\quad$ & 1.77$\quad$ & 0.31$\quad$  \\
\hline
\end{tabular}
\caption{BAO data sample (only angular) from \cite{deCarvalho:2017xye}.}
\label{tab-bao}
\end{table*}
With this sample, the theoretical BAO angular scale $\theta\left(z\right)$ can be computed using the angular diameter distance $d_{A}\left(z\right)$, which, for a flat universe is  
$\mathcal{D}\left(z\right)$,
$\mathcal{D}\left(z\right)=\frac{H_{0}}{c}\frac{r_{s}}{\theta\left(z\right)}\left(\frac{180}{\pi}\right)\,.$
where $r_{s}$ is the sound horizon of the primordial photon-baryon fluid .

\item[(d)] CMB from Planck 2018. We consider the only latest CMB constraint from the Planck Collaboration \cite{Aghanim:2018eyx} with $H_0 = 67.36 \pm 0.54 \text{km} \text{s}^{-1}\text{Mpc}^{-1}$. The CMB posterior can be computed by performing a MCMC modified chain using this $H_0$ value\footnote{\url{http://www.esa.int/Science_Exploration/Space_Science/Planck}}.

\item[(e)] Lyman-$\alpha$.
Large-scale structure measurements via the detection of Lyman-$\alpha$ (Ly$\alpha$) radiation in both emission and absorption. This sample provide constraints on the interactions dark matter-dark radiation through the Lyman-$\alpha$ forest flux power spectrum that comes from absorption lines in the spectra of distant quasars. We consider HIRES/MIKE data samples at $z:[4.2, 5.4]$ in 10 k-bins with 49 ($k,z$) data point (see \cite{Archidiacono:2019wdp} and references therein).

\end{itemize}

\section{Methodology using late-time samples}
\label{sec:method-late}

In this section we present the statistical results for each of the $f(R)$ models described by using our generic proposal  (\ref{eq:JJEeos}). To achieve this we include the C.L analysis for each free parameter space at 1-2-$\sigma$ level and their bes fits constrained by the late-time data samples (Pantheon SNeIa, CC and BAO) described in the last section.

\begin{table*}
	\begin{center}
		\caption{Interval priors used for the late-time samples (SNeIa + CC + BAO)}

		\begin{tabular}{|c|c|c|c|c|c|c|c|c|c|}
			\hline
			& \multicolumn{9}{ |c| }{\textbf{Priors}}  \\ \hline
			& \multicolumn{3}{ |c| }{\textbf{Starobinsky model}} & \multicolumn{3}{ |c| }{\textbf{Hu - Sawicki model}} & \multicolumn{3}{ |c| }{\textbf{Exponential model}} \\ \hline
			Parameter & mean & min & max & mean & min & max & mean & min & max \\ \hline
			$\omega_{b}$ & $2.2030$ & None & \text{None} & $2.2030$ & None & \text{None} & $2.2030$ & None & \text{None} \\
			$\omega_{cdm}$ & $0.106$ & None & None  & $0.106$ & None & None  & $0.106$ & None & None \\
			$\omega_{0}$ & $0.09$ & $-2$ & $2$ & $ 0.024$ & $-2$ & $2$ & $0.358$ & $-2$ & $2$ \\
			$M$ & $-19.02$ & None & None  & $-19.02$ & None & None  & $-19.02$ & None & None  \\
			$H_0$ & $67.8$ & None & None  & $67.8$ & None & None  & $67.8$ & None & None  \\
			\hline
		\end{tabular} \label{tab:prior_lateTime1}
	\end{center}
\end{table*}

\begin{table*}
	\begin{center}
		\caption{Uniform priors used for the time-time samples (SNeIa + CC + BAO). \textit{None} indicates definition as free interval. }
		\begin{tabular}{|c|c|c|c|}
			\hline
			& \multicolumn{1}{ |c| }{\textbf{Starobinsky model}} & \multicolumn{1}{ |c| }{\textbf{Hu - Sawicki model}} & \multicolumn{1}{ |c| }{\textbf{Exponential model}} \\ \hline
			Parameter & Priors (uniform) & Priors (uniform) & Priors (uniform)\\ \hline
			$\omega_{b}$ & [None, None] & [None, None]  & [None, None] \\
			$\omega_{cdm}$ & [None, None]  & [None, None] & [None, None]  \\
			$\omega_{0}$ & $[-2,2]$ & $[-2,2]$  & $[-2,2]$ \\
			$M$ & [None, None] & [None, None] & [None, None]   \\
			$H_0$ & [None, None]  & [None, None]  & [None, None]  \\
			\hline
		\end{tabular} \label{tab:prior_lateTime2}
	\end{center}
\end{table*}


\subsection{Starobinsky model}
\begin{figure*}[ht]
	\centering
	\includegraphics[width=0.75\textwidth]{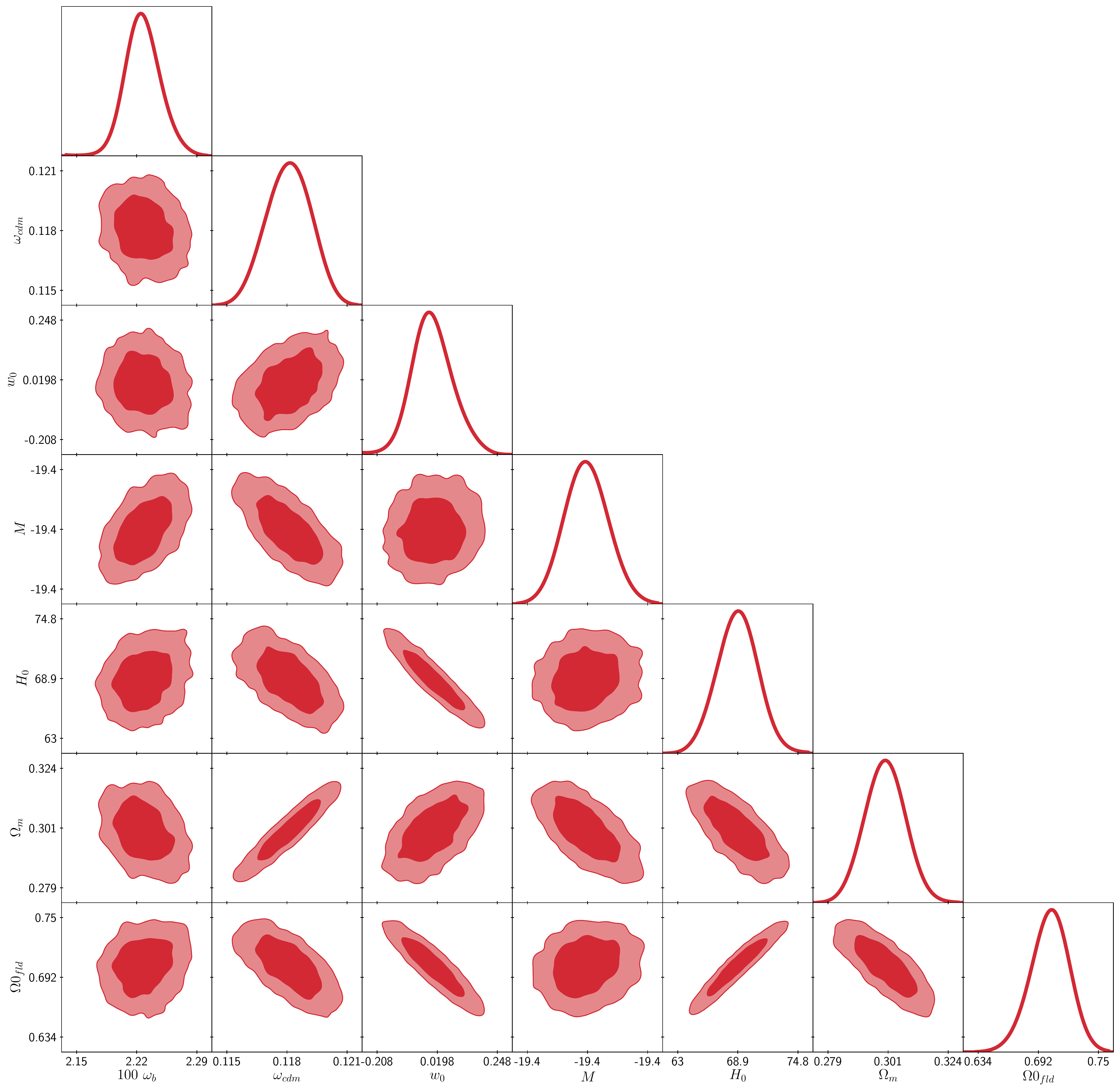}
	\caption{2-$\sigma$ C.L for the Starobinsky model using the late-time total sample (Pantheon SNeIa+ CC +BAO).}
	\label{fig:SpaceParameter_St_Ut}
\end{figure*}

For this model we obtain the Figure \ref{fig:SpaceParameter_St_Ut} with the bestfits reported in Table \ref{tab:late-Starobinsky}.

\begin{table*}
\centering
\caption{Best fits for the Starobinsky model using using the late-time total sample (Pantheon SNeIa+ CC +BAO). From the final fit we obtain: $-\ln{\cal L}_\mathrm{min} =523.46$ and $\chi^2=1047$.}
\begin{tabular}{|l|c|c|c|c|}
 \hline
Parameter & Best-fit & mean$\pm\sigma$ & 95\% lower & 95\% upper \\ \hline
$100~\omega_{b }$ &$2.225$ & $2.224_{-0.024}^{+0.021}$ & $2.18$ & $2.269$ \\
$\omega_{cdm }$ &$0.1182$ & $0.1182_{-0.0013}^{+0.0012}$ & $0.1158$ & $0.1207$ \\
$w_{0 }$ &$0.004789$ & $0.002704_{-0.082}^{+0.078}$ & $-0.154$ & $0.164$ \\
$M$ &$-19.4$ & $-19.4_{-0.014}^{+0.012}$ & $-19.42$ & $-19.37$ \\
$H_{0 }$ &$68.75$ & $68.85_{-2}^{+2.1}$ & $64.75$ & $72.91$ \\
$\Omega_{m }$ &$0.3005$ & $0.3004_{-0.0078}^{+0.0077}$ & $0.285$ & $0.3155$ \\
$\Omega_0$ &$0.7026$ & $0.7028_{-0.017}^{+0.021}$ & $0.6641$ & $0.7403$ \\
\hline
 \end{tabular}\label{tab:late-Starobinsky}
 \end{table*}

\subsection{Hu-Sawicki model}

\begin{figure*}[ht]
	\centering
	\includegraphics[width=0.75\textwidth]{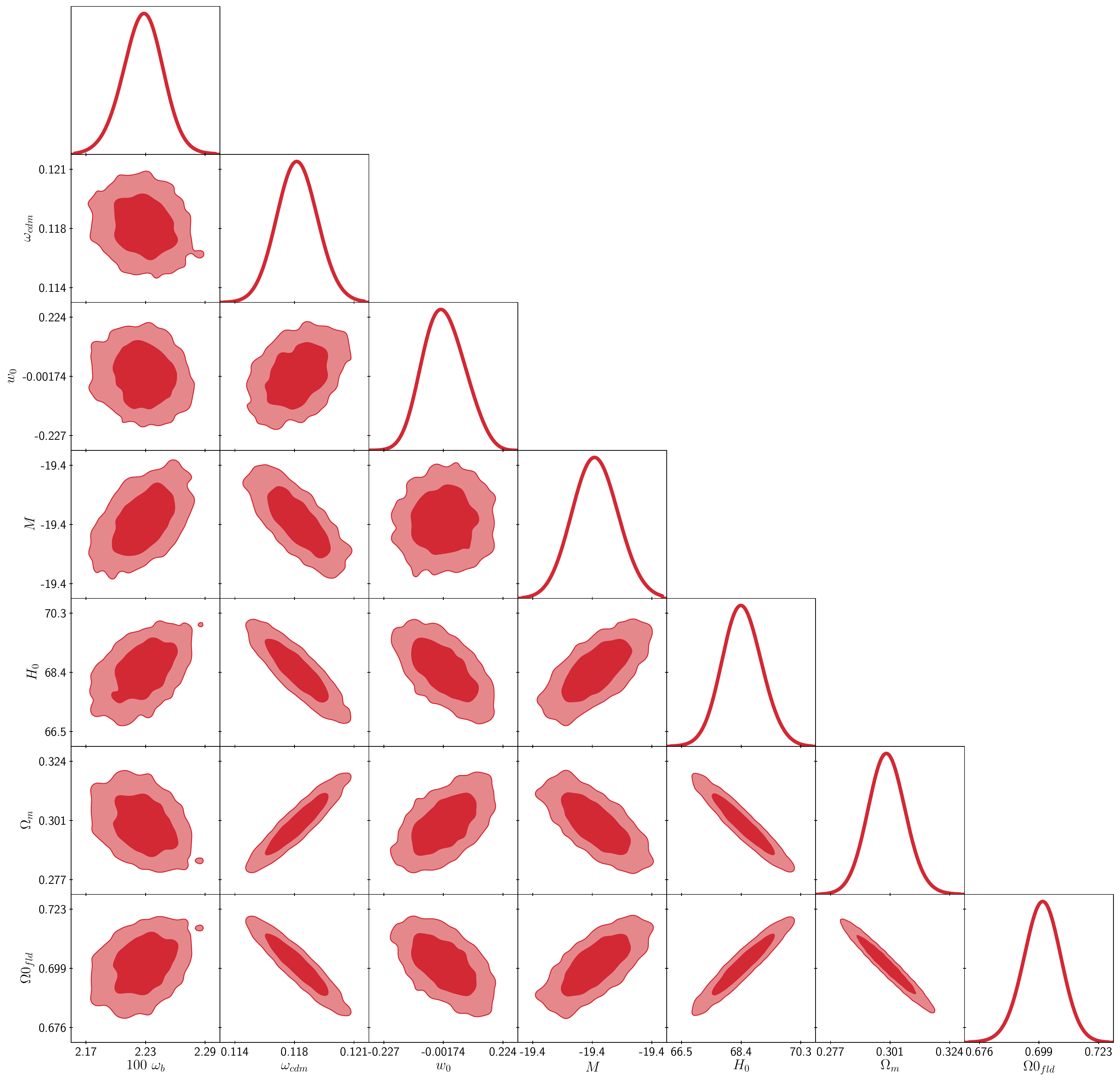}
	\caption{2-$\sigma$ C.L for the Hu-Sawicki using the late-time total sample (Pantheon SNeIa+ CC +BAO).}
	\label{fig:SpaceParameter_HS_Ut}
\end{figure*}

For this model we obtain the Figure \ref{fig:SpaceParameter_HS_Ut} with the bestfits reported in Table \ref{tab:H-S}.

\begin{table*}
\centering
\caption{Best fits for the Hu-Sawicki model using the late-time total sample (Pantheon SNeIa+ CC +BAO). From the final fit we obtain: $-\ln{\cal L}_\mathrm{min} =523.47$ and $\chi^2=1047$.}

\begin{tabular}{|l|c|c|c|c|}
 \hline
Parameter & Best-fit & mean$\pm\sigma$ & 95\% lower & 95\% upper \\ \hline
$100~\omega_{b }$ &$2.228$ & $2.227_{-0.023}^{+0.022}$ & $2.181$ & $2.272$ \\
$\omega_{cdm }$ &$0.1182$ & $0.118_{-0.0012}^{+0.0013}$ & $0.1155$ & $0.1205$ \\
$w_{0 }$ &$0.00793$ & $0.002035_{-0.088}^{+0.079}$ & $-0.1489$ & $0.1641$ \\
$M$ &$-19.4$ & $-19.4_{-0.014}^{+0.013}$ & $-19.42$ & $-19.37$ \\
$H_{0 }$ &$68.41$ & $68.47_{-0.64}^{+0.69}$ & $67.09$ & $69.7$ \\
$\Omega_{m }$ &$0.3001$ & $0.2994_{-0.0083}^{+0.008}$ & $0.2834$ & $0.3154$ \\
$\Omega_0$ &$0.6998$ & $0.7005_{-0.008}^{+0.0083}$ & $0.6845$ & $0.7165$ \\
\hline
 \end{tabular} \label{tab:H-S}
 \end{table*}

\subsection{Exponential model}

\begin{figure*}[ht]
	\centering
	\includegraphics[width=0.75\textwidth]{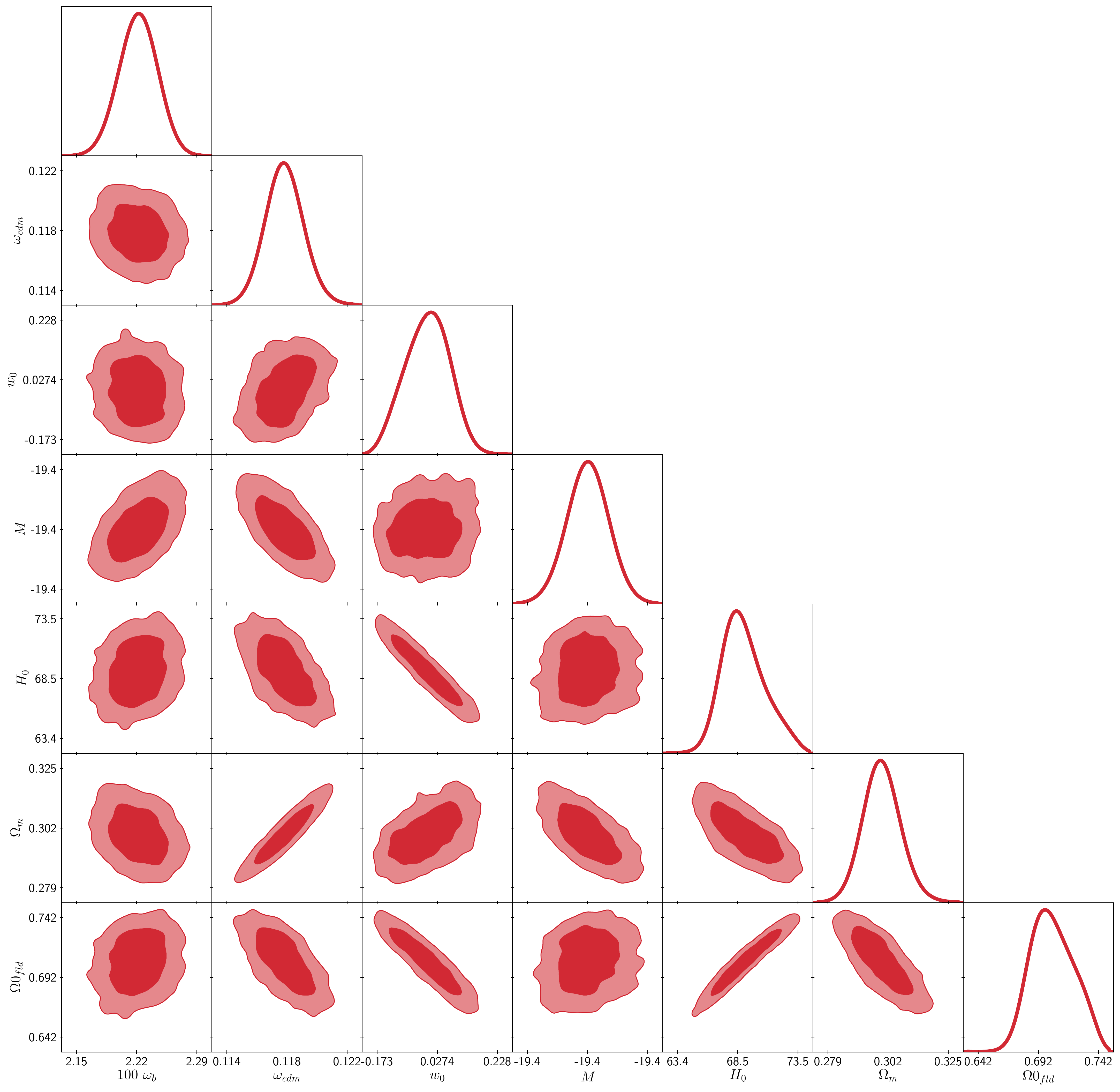}
	\caption{2-$\sigma$ C.L for the Exponential model using the late-time total sample (Pantheon SNeIa+ CC +BAO).}
	\label{fig:SpaceParameter_Exp_Ut}
\end{figure*}

For this model we obtain the Figure \ref{fig:SpaceParameter_Exp_Ut} with the bestfits reported in Table \ref{tab:exp}.

\begin{table*}
\centering
\caption{Best fits for the Exponential model using the late-time total sample (Pantheon SNeIa+ CC +BAO). From the final fit we obtain: $-\ln{\cal L}_\mathrm{min} =523.463$ and $\chi^2=1047$.}
\begin{tabular}{|l|c|c|c|c|}
 \hline
Parameter & Best-fit & mean$\pm\sigma$ & 95\% lower & 95\% upper \\ \hline
$100~\omega_{b }$ &$2.224$ & $2.226_{-0.024}^{+0.022}$ & $2.181$ & $2.271$ \\
$\omega_{cdm }$ &$0.1182$ & $0.1181_{-0.0013}^{+0.0012}$ & $0.1156$ & $0.1206$ \\
$w_{0 }$ &$0.0001941$ & $-0.006569_{-0.082}^{+0.08}$ & $-0.1559$ & $0.1364$ \\
$M$ &$-19.4$ & $-19.4_{-0.013}^{+0.012}$ & $-19.42$ & $-19.37$ \\
$H_{0 }$ &$68.86$ & $69.11_{-2.2}^{+1.9}$ & $65.44$ & $72.95$ \\
$\Omega_{m }$ &$0.3001$ & $0.2995_{-0.0081}^{+0.0072}$ & $0.2844$ & $0.3147$ \\
$\Omega_0$ &$0.7038$ & $0.7053_{-0.017}^{+0.021}$ & $0.6706$ & $0.7407$ \\
\hline
 \end{tabular} \label{tab:exp}
 \end{table*}

\section{Methodology using early-time samples}
\label{sec:method-early}

In this section we present the statistical results for each of the $f(R)$ models described by using our generic proposal  (\ref{eq:JJEeos}). To achieve this we include the C.L analysis for each free parameter space at 1-2-$\sigma$ level and their bes fits constrained by the early-time data samples (Planck 2018 and Ly-$\alpha$) described in the last section.

\begin{table*}
	\begin{center}
		\caption{Interval priors used for the early-time samples (Planck2018 + Ly-$\alpha$).
		}
		\begin{tabular}{|c|c|c|c|c|c|c|c|c|c|}
			\hline
			& \multicolumn{9}{ |c| }{\textbf{Priors}}  \\ \hline
			& \multicolumn{3}{ |c| }{\textbf{Starobinsky model}} & \multicolumn{3}{ |c| }{\textbf{Hu - Sawicki model}} & \multicolumn{3}{ |c| }{\textbf{Exponential model}} \\ \hline
			Parameter & mean & min & max & mean & min & max & mean & min & max \\ \hline
			$\omega_{b}$ & $2.2377$ & None & \text{None} & $2.2377$ & None & \text{None} & $2.2377$ & None & \text{None} \\
			$\omega_{cdm}$ & $0.12010$ & None & None  & $0.12010$ & None & None  & $0.12010$ & None & None \\
			$\omega_{0}$ & $0.09$ & $-2$ & $2$ & $0.024$ & $-2$ & $2$ & $ 0.358$ & $-2$ & $2$ \\
			$H_0$ & $67.8$ & 55 & 80 & $67.8$ & 55 & 80  & $67.8$ & 55 & 80  \\
			$ln10^{10}A_{s}$ & $3.0447$ & None & None  & $3.0447$ & None & None  & $3.0447$ & None & None  \\
			$n_{s}$ & $0.9659$ & None & None  & $0.9659$ & None & None  & $0.9659$ & None & None  \\
			$\tau_{reio }$ & $0.0543$ & $0.004$ & None  & $0.0543$ & $0.004$ & None  & $0.0543$ & $0.004$ & None  \\
			\hline
		\end{tabular} \label{tab:prior_earlyTime}
	\end{center}
\end{table*}

\begin{table*}
	\begin{center}
		\caption{Uniform priors used for the early-time samples (Planck2018 + Ly-$\alpha$).   
		}
		\begin{tabular}{|c|c|c|c|}
			\hline
			& \multicolumn{1}{ |c| }{\textbf{Starobinsky model}} & \multicolumn{1}{ |c| }{\textbf{Hu - Sawicki model}} & \multicolumn{1}{ |c| }{\textbf{Exponential model}} \\ \hline
			Parameter & Priors (uniform) & Priors (uniform) & Priors (uniform)\\ \hline
			$\omega_{b}$ & [None, None] & [None, None]  & [None, None] \\
			$\omega_{cdm}$ & [None, None]  & [None, None] & [None, None]  \\
			$\omega_{0}$ & $[-2,2]$ & $[-2,2]$  & $[-2,2]$ \\
			$H_0$ & $[55, 80]$  & $[55, 80]$  & $[55, 80]$  \\
			$ln10^{10}A_{s}$ & [None, None] & [None, None] & [None, None]    \\
			$n_{s}$ & [None, None] & [None, None] & [None, None]    \\
			$\tau_{reio }$ & $[0.004$, None] & $[0.004$, None] & $[0.004$, None]   \\
			\hline
		\end{tabular} \label{tab:prior_earlyTime2}
	\end{center}
\end{table*}

\subsection{Starobinsky model}

\begin{figure*}[ht]
	\centering
	\includegraphics[width=0.75\textwidth]{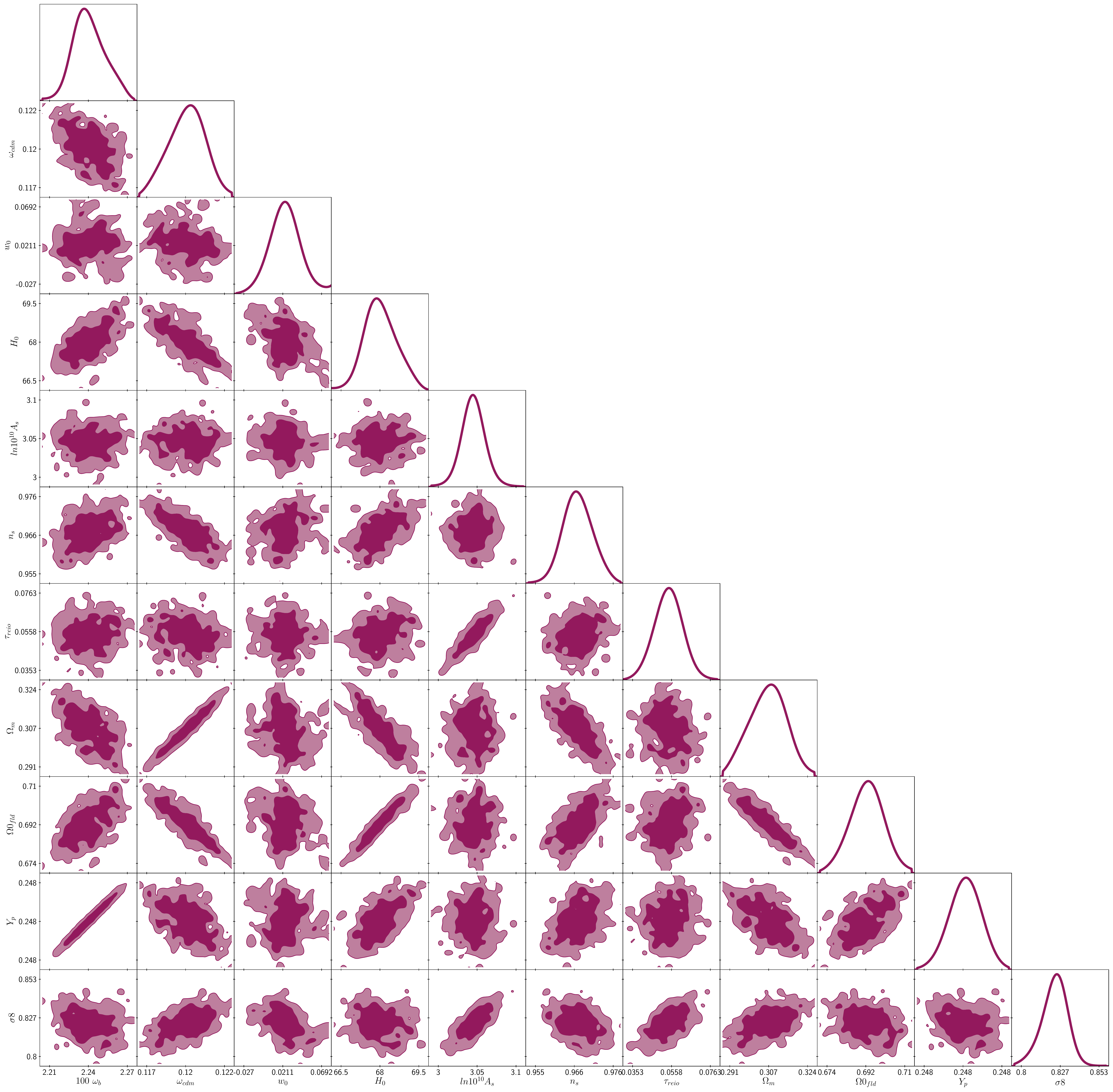}
	\caption{2-$\sigma$ C.L for the Starobinsky model using the early-time total sample (Planck 2018+Ly$\alpha$).}
	\label{fig:SpaceParameter_St_Et}
\end{figure*}

\begin{figure*}[ht]
	\centering
	\includegraphics[width=0.45\textwidth]{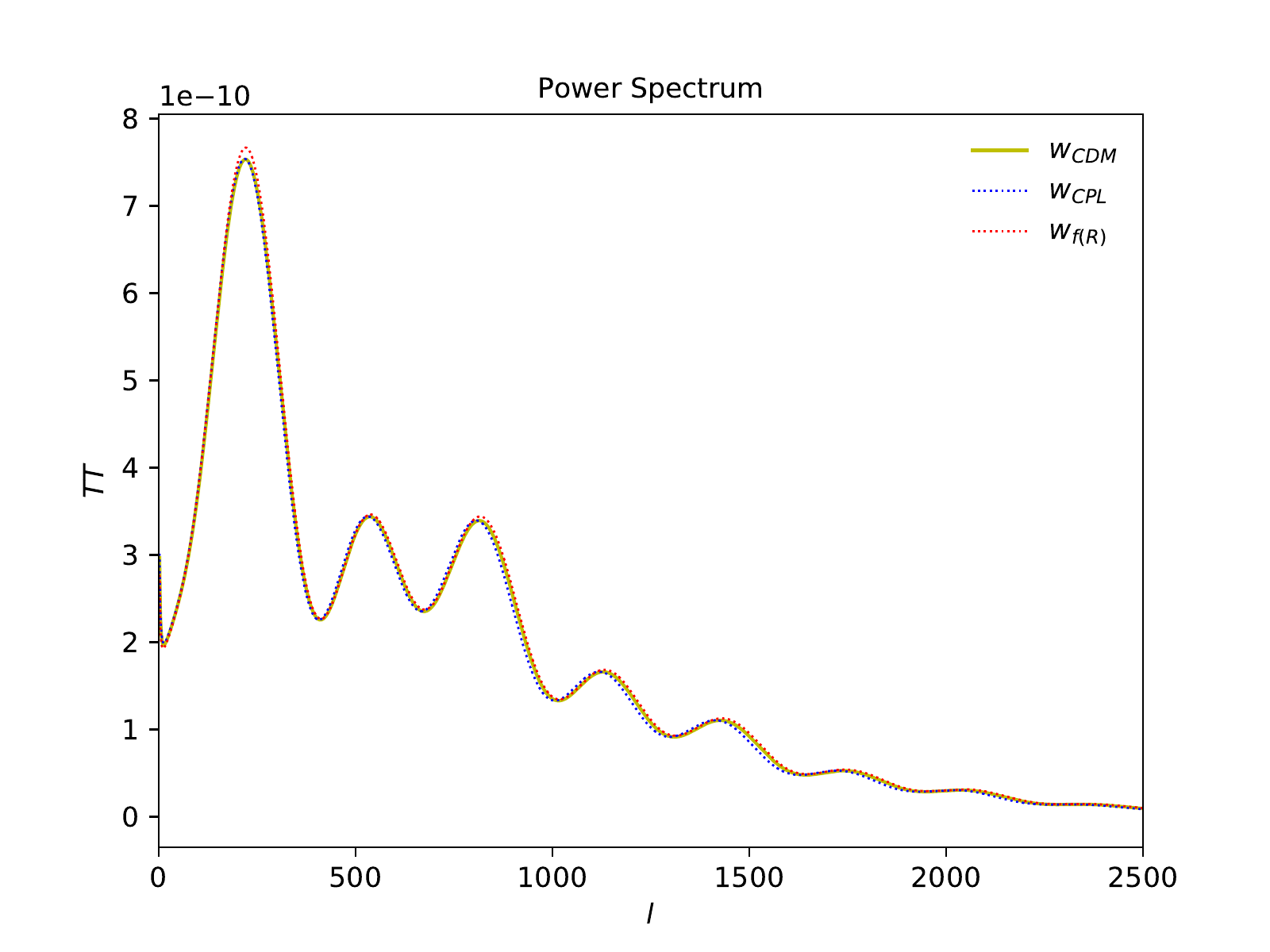}
	\includegraphics[width=0.45\textwidth]{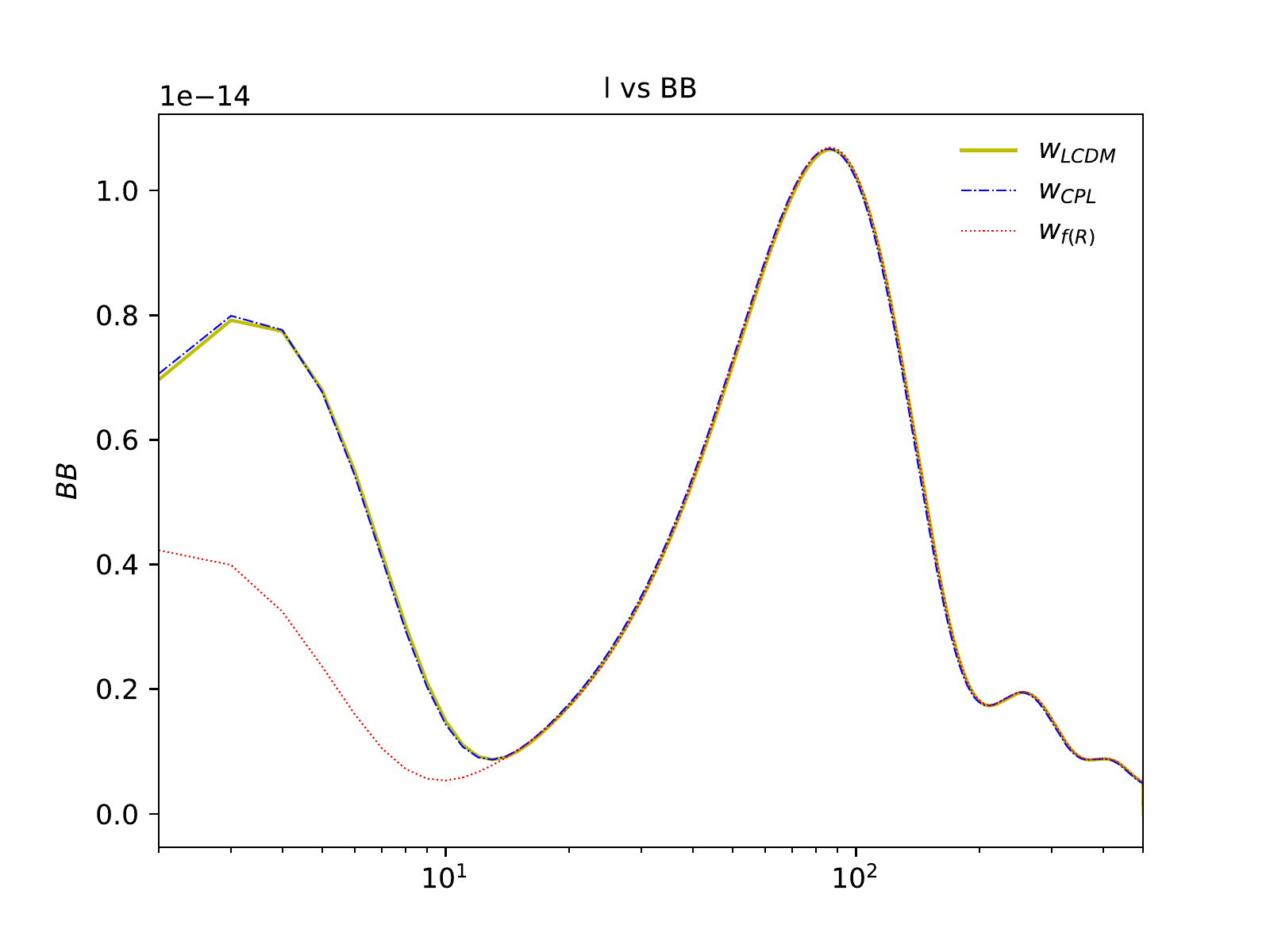}
	\includegraphics[width=0.45\textwidth]{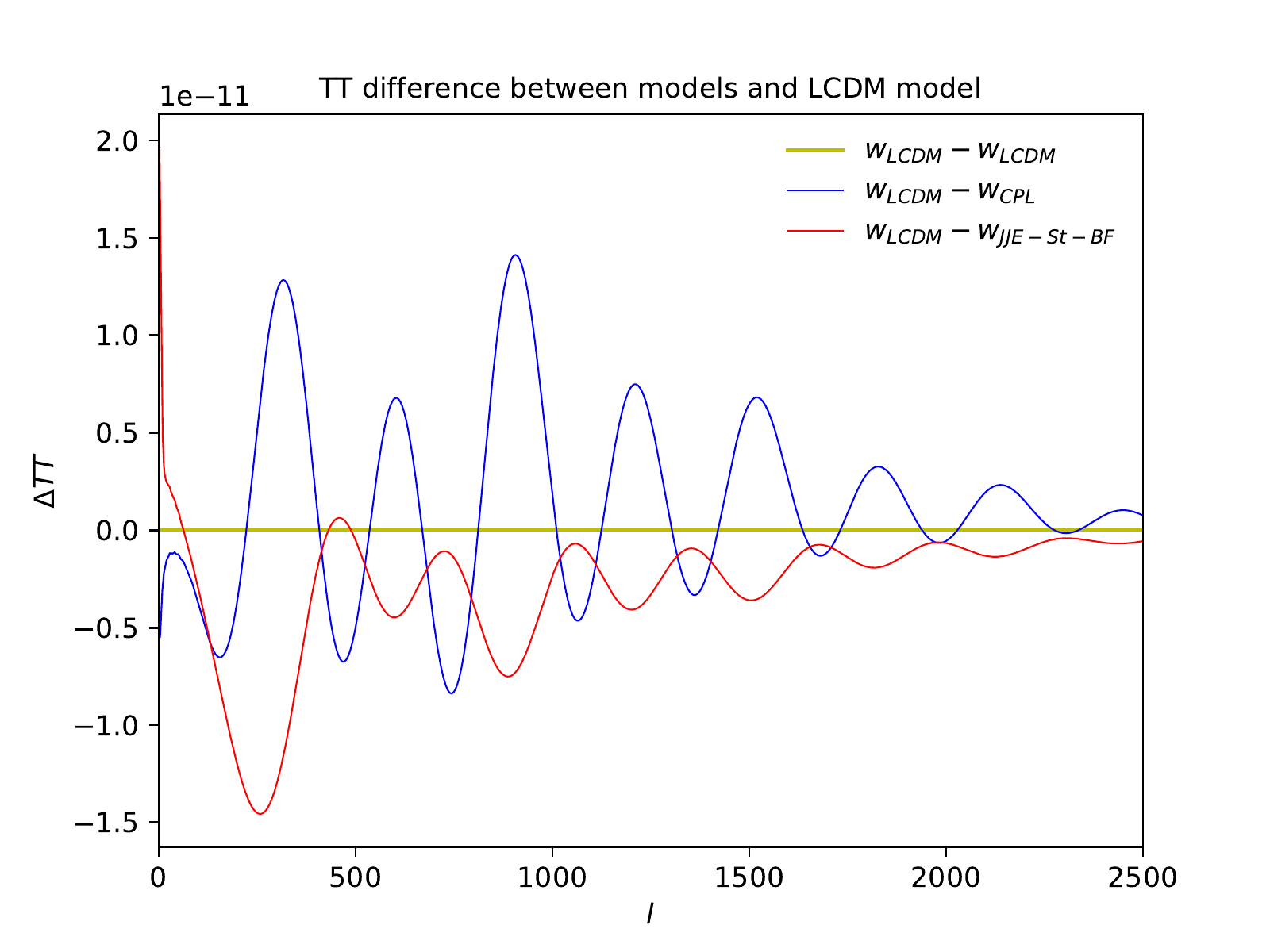}
	\includegraphics[width=0.45\textwidth]{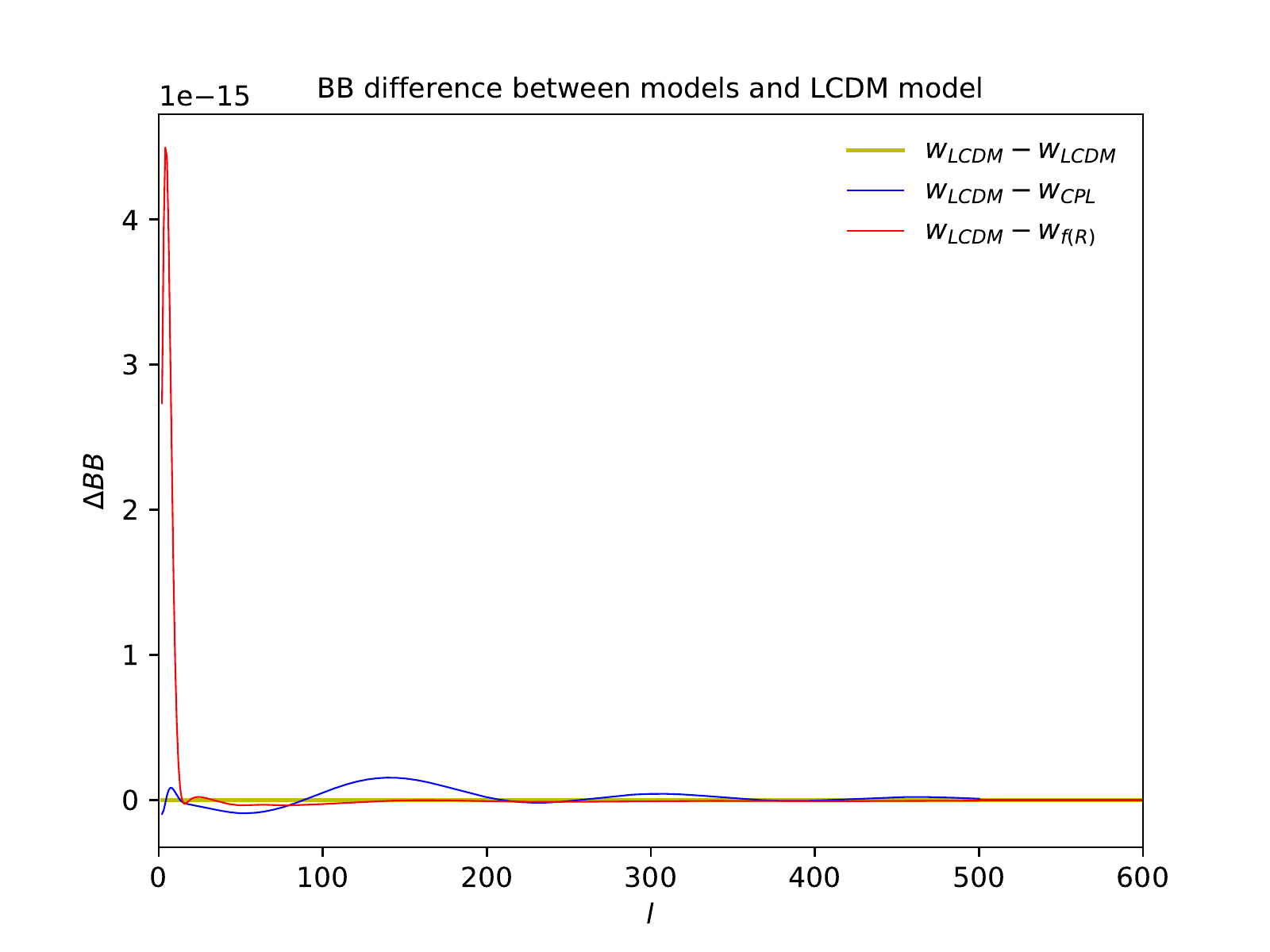}
	\caption{Power spectra for Starobinsky model derived from (\ref{eq:JJEeos}) and using the best-fits of the parameters obtained with the total sample (Pantheon SNeIa + BAO + CC + Planck 2018 + Ly-$\alpha$) reported in Table \ref{tab:total-Starobinsky}. We compare our parameterisation in this model with standard ones: $\Lambda$CDM and $CPL$ as $w(z) =w_0 +\frac{w_1 z}{ 1+z}$. \textit{Top (from Left to Right)}: TT and BB power spectrum. \textit{Bottom (from Left to Right)}: $\Delta$TT and $\Delta$BB ratio between standard models.}
	\label{fig:PS_Starobinsky}
\end{figure*}

For this model we obtain the C.L analysis in Figure \ref{fig:SpaceParameter_St_Et} with the bestfits reported in Table \ref{tab:sta-early} and their power spectra in Figure \ref{fig:PS_Starobinsky}

\begin{table*}
\centering
\caption{Best fits for the Starobinsky model using the early-time total sample (Planck 2018+Ly$\alpha$). From the final fit we obtain: $-\ln{\cal L}_\mathrm{min} =1406.54$ and $\chi^2=2813$.}
\begin{tabular}{|l|c|c|c|c|}
	\hline
	Parameter & Best-fit & mean$\pm\sigma$ & 95\% lower & 95\% upper \\ \hline
	$100~\omega_{b }$ &$2.256$ & $2.24_{-0.0138644}^{+-0.013}$ & $2.242$ & $2.269$ \\
	$\omega_{cdm }$ &$0.1192$ & $0.1196_{-1.345e-03}^{+-1.345e-03}$ & $0.118$ & $0.120$ \\
	$w_{0 }$ &$0.05334$ & $0.02378_{-0.018}^{+0.018}$ & $0.035$ & $0.071$ \\
	$H_{0 }$ &$67.72$ & $68.02_{-0.645}^{+0.645}$ & $67.074$ & $68.365$ \\
	$ln10^{10}A_{s }$ &$3.043$ & $3.043_{-0.018}^{+0.013}$ & $3.01$ & $3.073$ \\
	$n_{s }$ &$0.971$ & $0.9667_{-4.205e-03}^{+4.205e-03}$ & $0.966$ & $0.975$ \\
	$\tau_{reio }$ &$0.05718$ & $0.05409_{-0.0068}^{+0.0081}$ & $0.050$ & $0.064$ \\
	$\Omega_{m }$ &$0.3041$ & $0.3071_{-0.034}^{+0.034}$ & $0.269$ & $0.338$ \\
	$\Omega0_{fld }$ &$0.6907$ & $0.6928_{-0.0083}^{+0.0086}$ & $0.678$ & $0.710$ \\
	$Y_{p }$ &$0.2479$ & $0.2479_{-6.2e-05}^{+6.3e-05}$ & $0.247$ & $0.248$ \\
	$\sigma8$ &$0.8174$ & $0.8227_{-0.0076}^{+0.0098}$ & $0.804$ & $0.84$ \\
	\hline
\end{tabular} \label{tab:sta-early}
 \end{table*}

\subsection{Hu-Sawicky model}

\begin{figure*}[ht]
	\centering
	\includegraphics[width=0.75\textwidth]{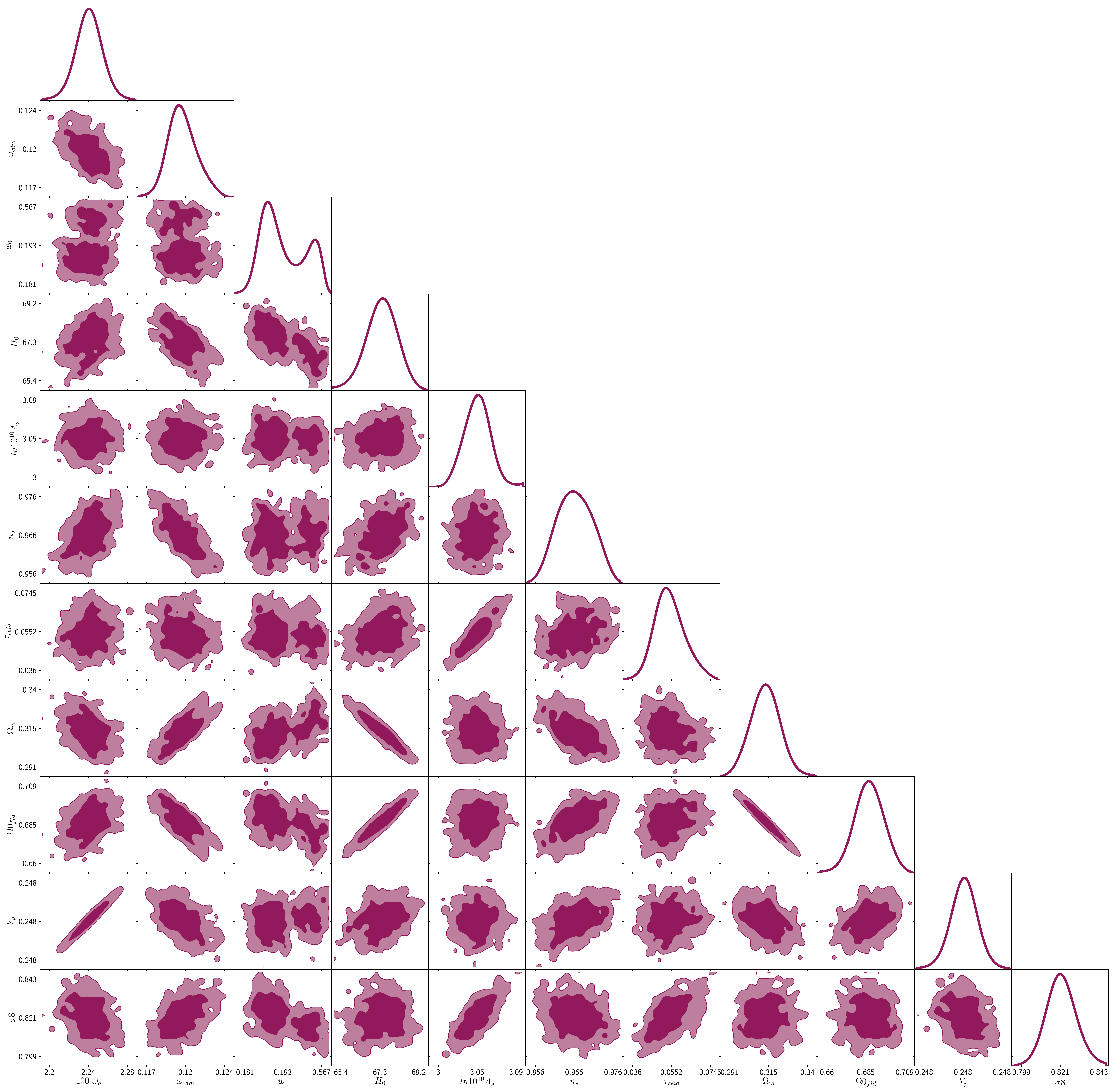}
	\caption{2-$\sigma$ C.L for the Hu-Sawicky  model using the early-time total sample (Planck 2018+Ly$\alpha$).
		}
	\label{fig:SpaceParameter_HS_Et}
\end{figure*}

\begin{figure*}[ht]
	\centering
	\includegraphics[width=0.45\textwidth]{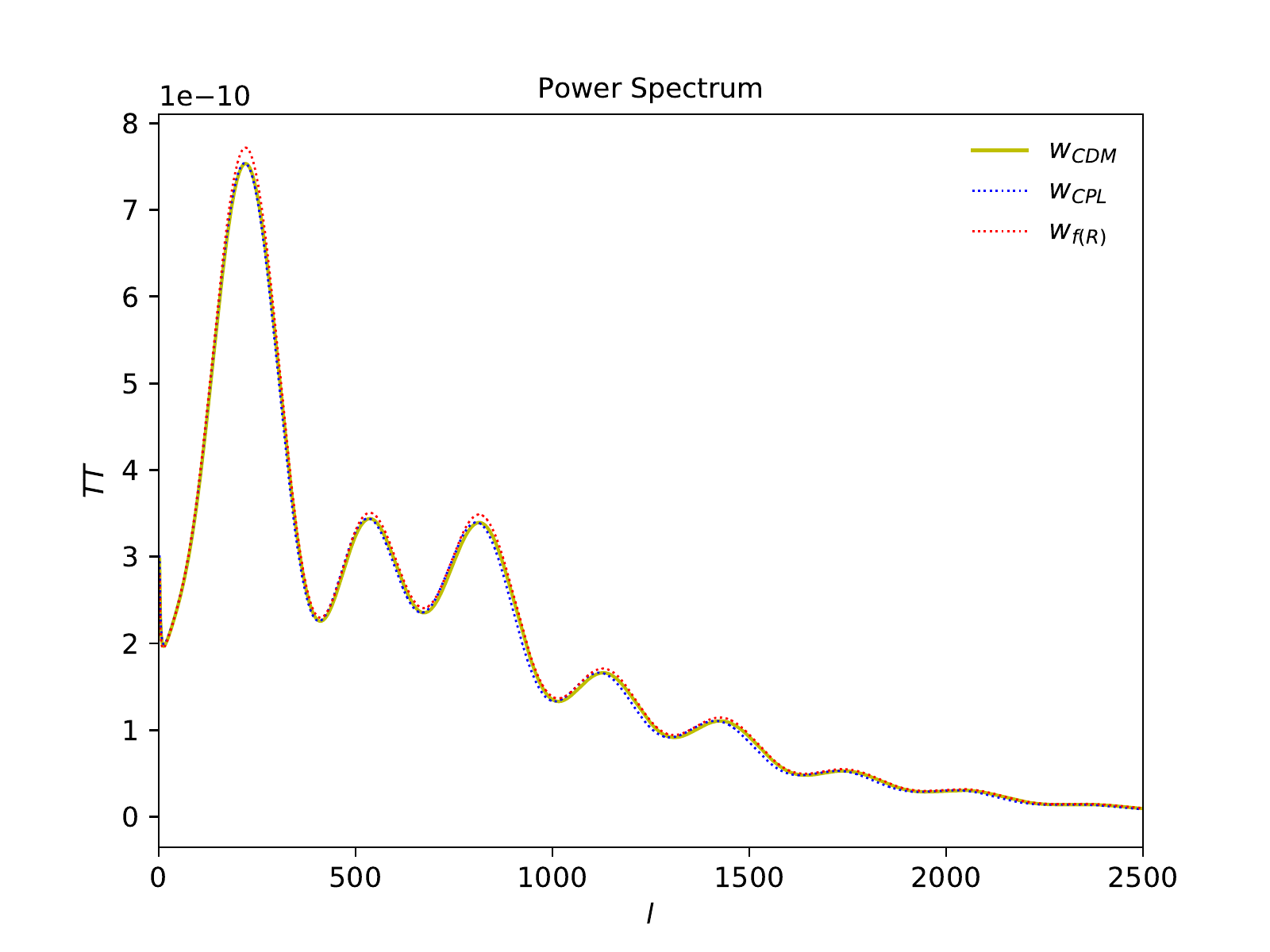}
	\includegraphics[width=0.45\textwidth]{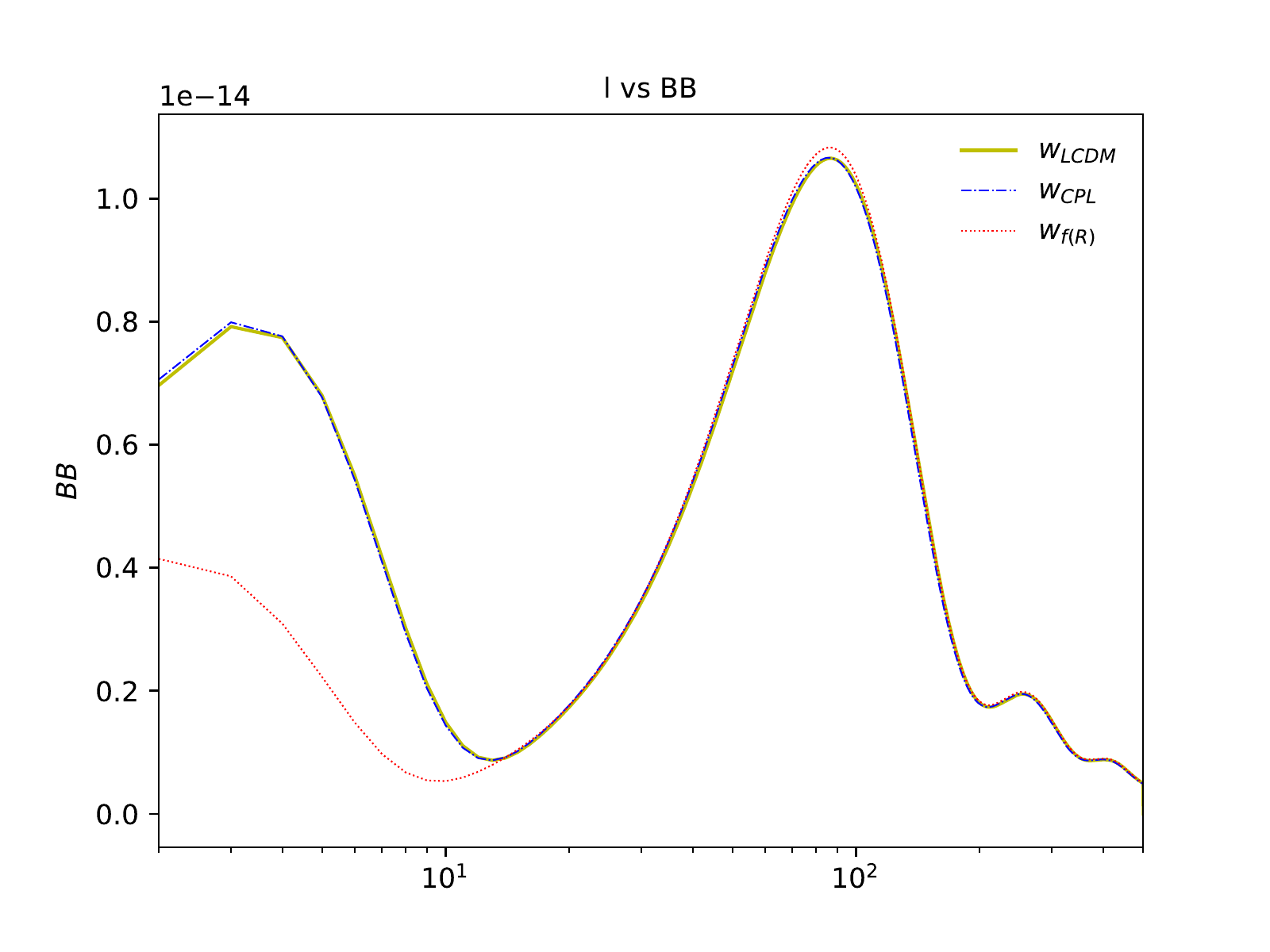}
	\includegraphics[width=0.45\textwidth]{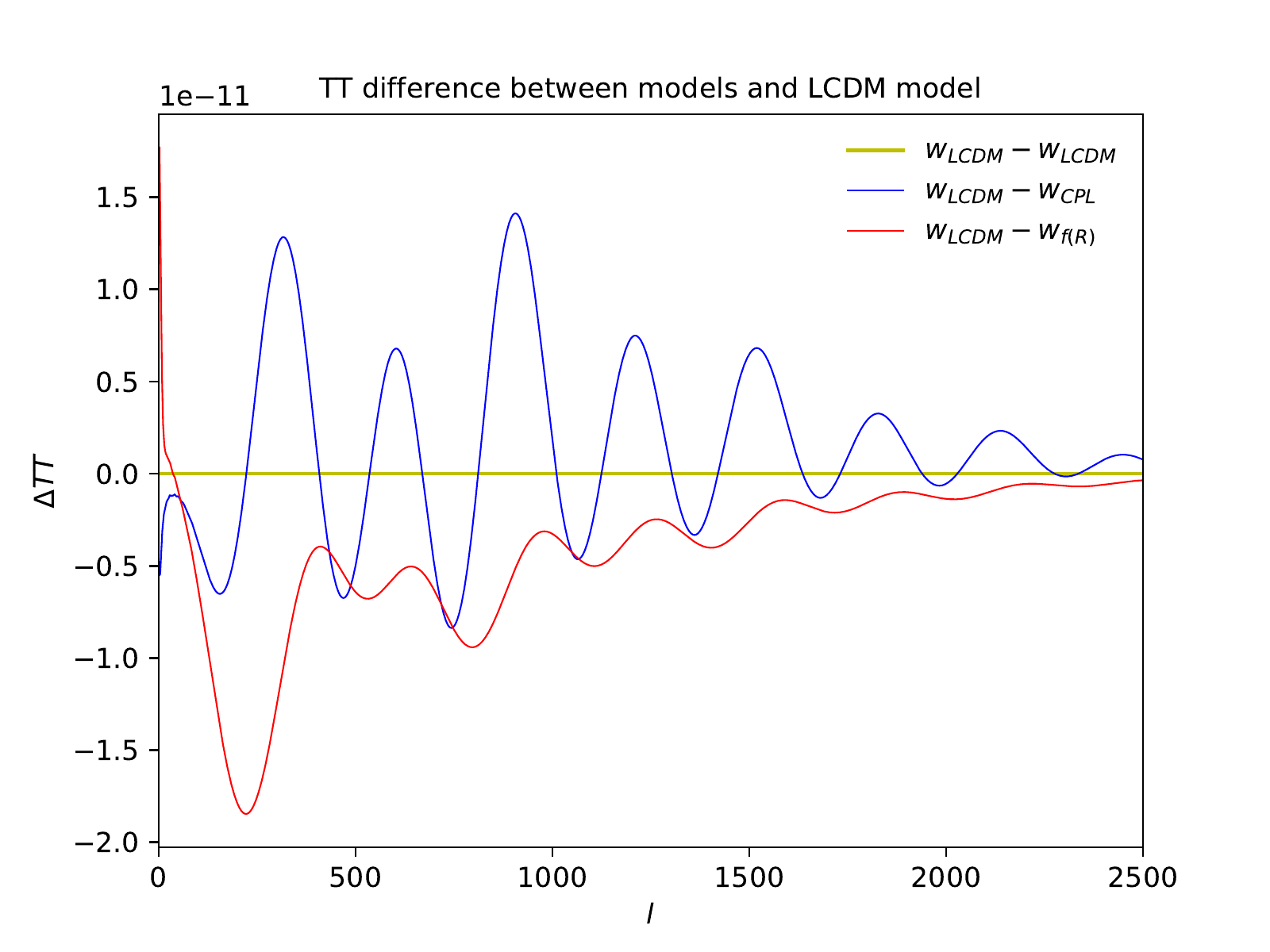}
	\includegraphics[width=0.45\textwidth]{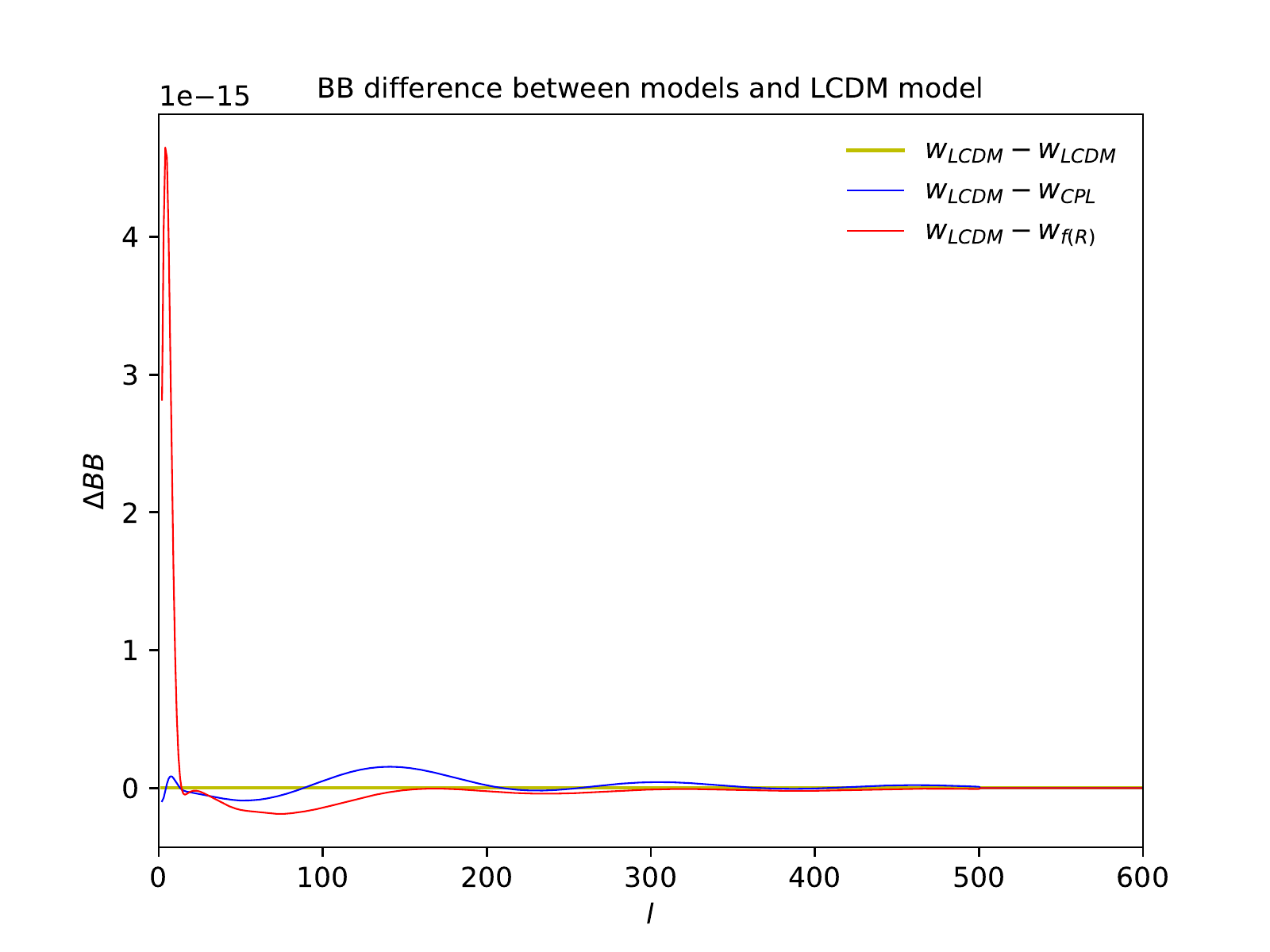}
	\caption{Power spectra for Hu-Sawicki model derived from (\ref{eq:JJEeos}) and using the best-fits of the parameters obtained with the total sample (Pantheon SNeIa + BAO + CC + Planck 2018 + Ly-$\alpha$) reported in Table \ref{tab:total-exp}. We compare our parameterisation in this model with standard ones: $\Lambda$CDM and $CPL$ as $w(z) =w_0 +\frac{w_1 z}{ 1+z}$. \textit{Top (from Left to Right)}: TT and BB power spectrum. \textit{Bottom (from Left to Right)}: $\Delta$TT and $\Delta$BB ratio between standard models.}
	\label{fig:PS_HuSawicki}
\end{figure*}

For this model we obtain the C.L analysis in Figure \ref{fig:SpaceParameter_HS_Et} with the bestfits reported in Table \ref{tab:HS-early} and their power spectra in Figure \ref{fig:PS_HuSawicki}

\begin{table*}
\centering
\caption{Best fits for the Hu-Sawicki model  using the early-time total sample (Planck 2018+Ly$\alpha$). From the final fit we obtain: $-\ln{\cal L}_\mathrm{min} =1403.58$ and $\chi^2=2807$.}
\begin{tabular}{|l|c|c|c|c|}
	\hline
	Parameter & Best-fit & mean$\pm\sigma$ & 95\% lower & 95\% upper \\ \hline
	$100~\omega_{b }$ &$2.229$ & $2.242_{-0.013}^{+0.016}$ & $2.213$ & $2.271$ \\
	$\omega_{cdm }$ &$0.12$ & $0.1199_{-0.0015}^{+0.0013}$ & $0.119$ & $0.121$ \\
	$w_{0 }$ &$0.07773$ & $0.2574_{-0.289}^{+0.289}$ & $-0.212$ & $0.367$ \\
	$H_{0 }$ &$67.6$ & $67.2_{-0.68}^{+1.2}$ & $66.92$ & $68.8$ \\
	$ln10^{10}A_{s }$ &$3.021$ & $3.044_{-0.017}^{+0.016}$ & $3.009$ & $3.077$ \\
	$n_{s }$ &$0.9657$ & $0.9663_{-0.004}^{+0.004}$ & $0.961$ & $0.970$ \\
	$\tau_{reio }$ &$0.04598$ & $0.05366_{-0.0091}^{+0.0068}$ & $0.03711$ & $0.07158$ \\
	$\Omega_{m }$ &$0.3114$ & $0.3153_{-0.013}^{+0.0089}$ & $0.298$ & $0.320$ \\
	$\Omega0_{fld }$ &$0.6885$ & $0.6846_{-0.0089}^{+0.013}$ & $0.6796$ & $0.701$ \\
	$Y_{p }$ &$0.2478$ & $0.2479_{-5.8e-05}^{+6.6e-05}$ & $0.2477$ & $0.248$ \\
	$\sigma8$ &$0.8144$ & $0.8193_{-0.0095}^{+0.01}$ & $0.805$ & $0.824$ \\
	\hline
\end{tabular} \label{tab:HS-early}
 \end{table*}

\subsection{Exponential model}

\begin{figure*}[ht]
	\centering
	\includegraphics[width=0.75\textwidth]{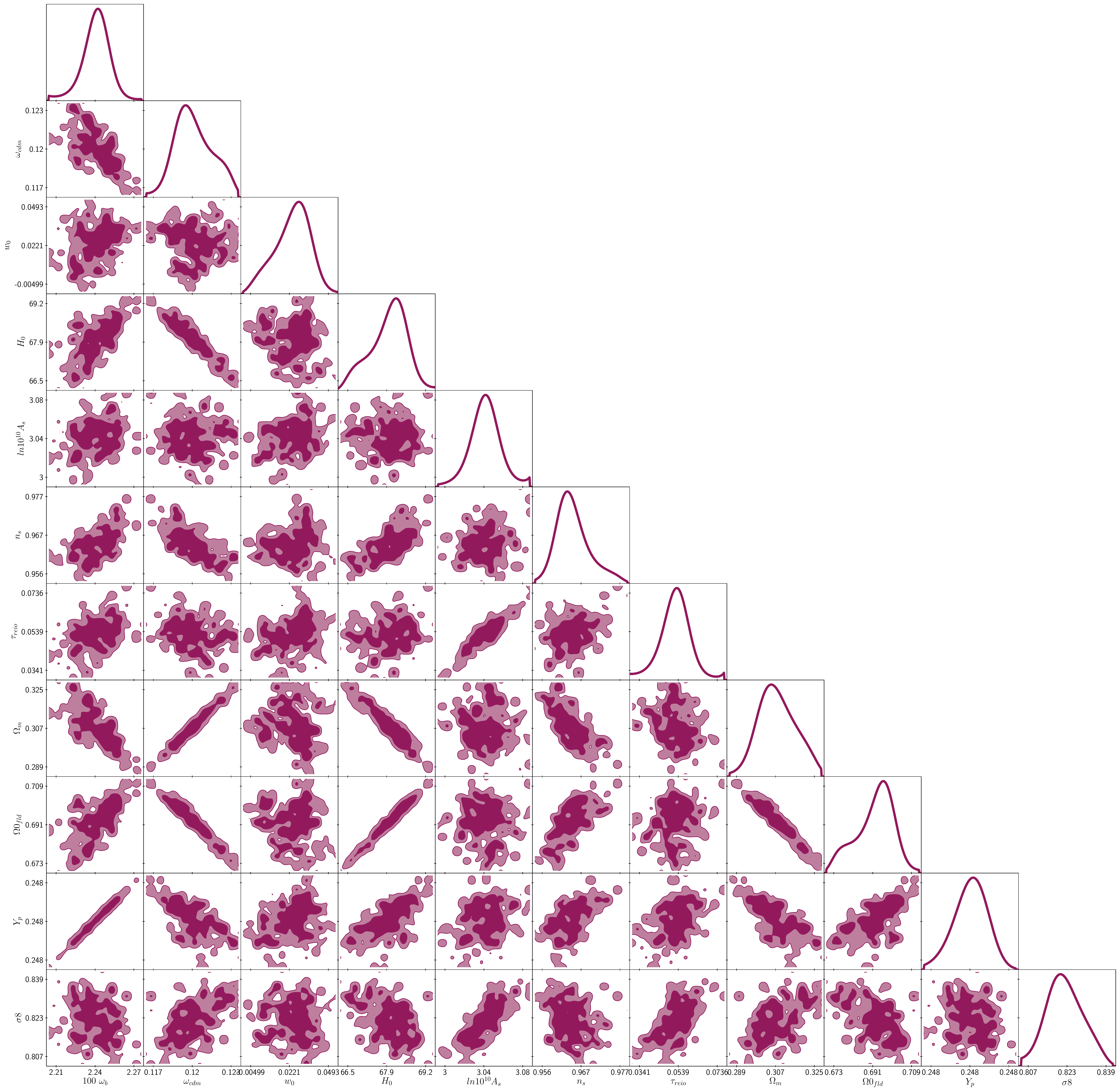}
	\caption{2-$\sigma$ C.L for the Exponential model using the early-time total sample (Planck 2018+Ly$\alpha$).}
	\label{fig:SpaceParameter_Exp_Et}
\end{figure*}

\begin{figure*}[ht]
	\centering
	\includegraphics[width=0.45\textwidth]{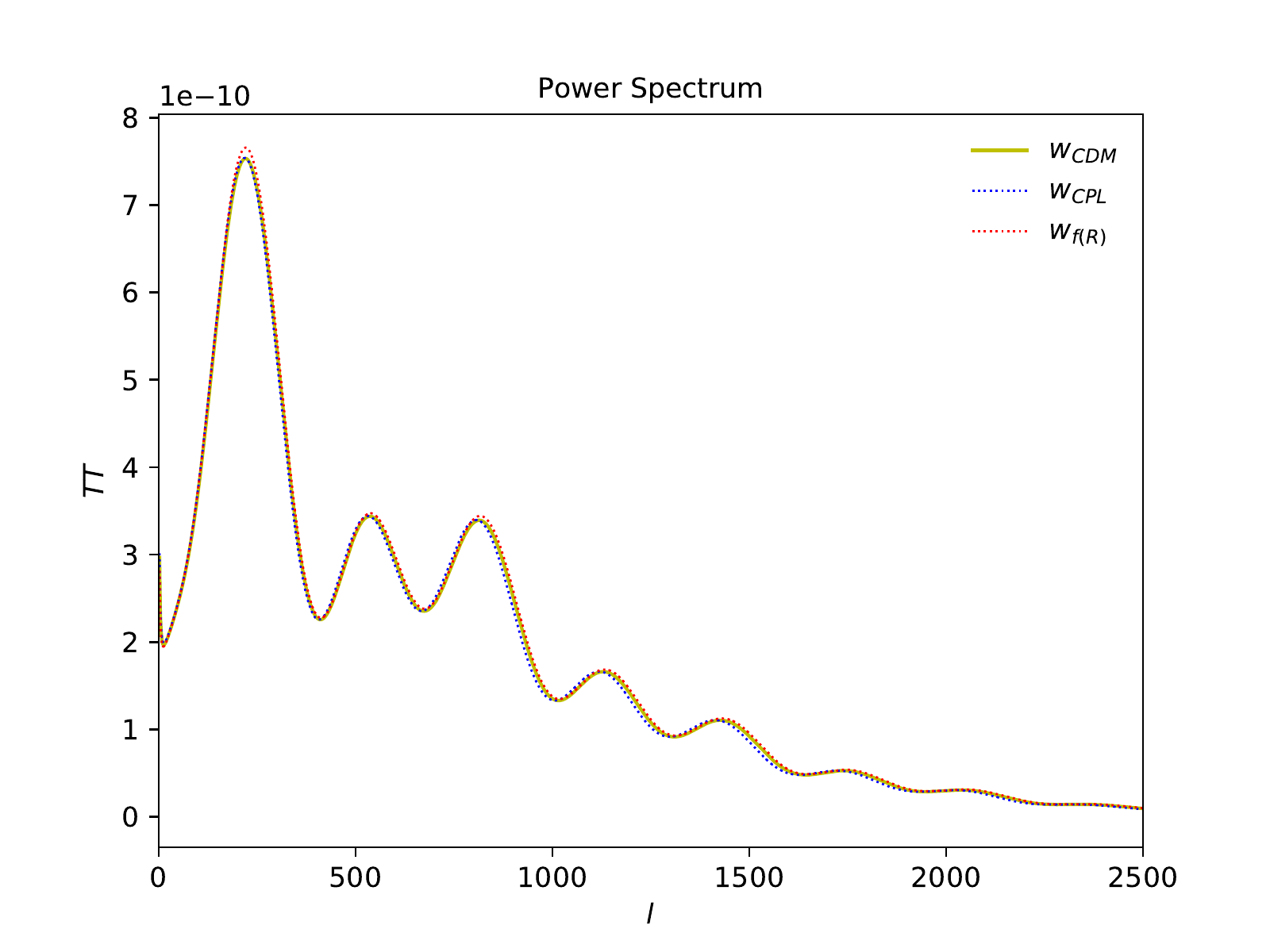}
	\includegraphics[width=0.45\textwidth]{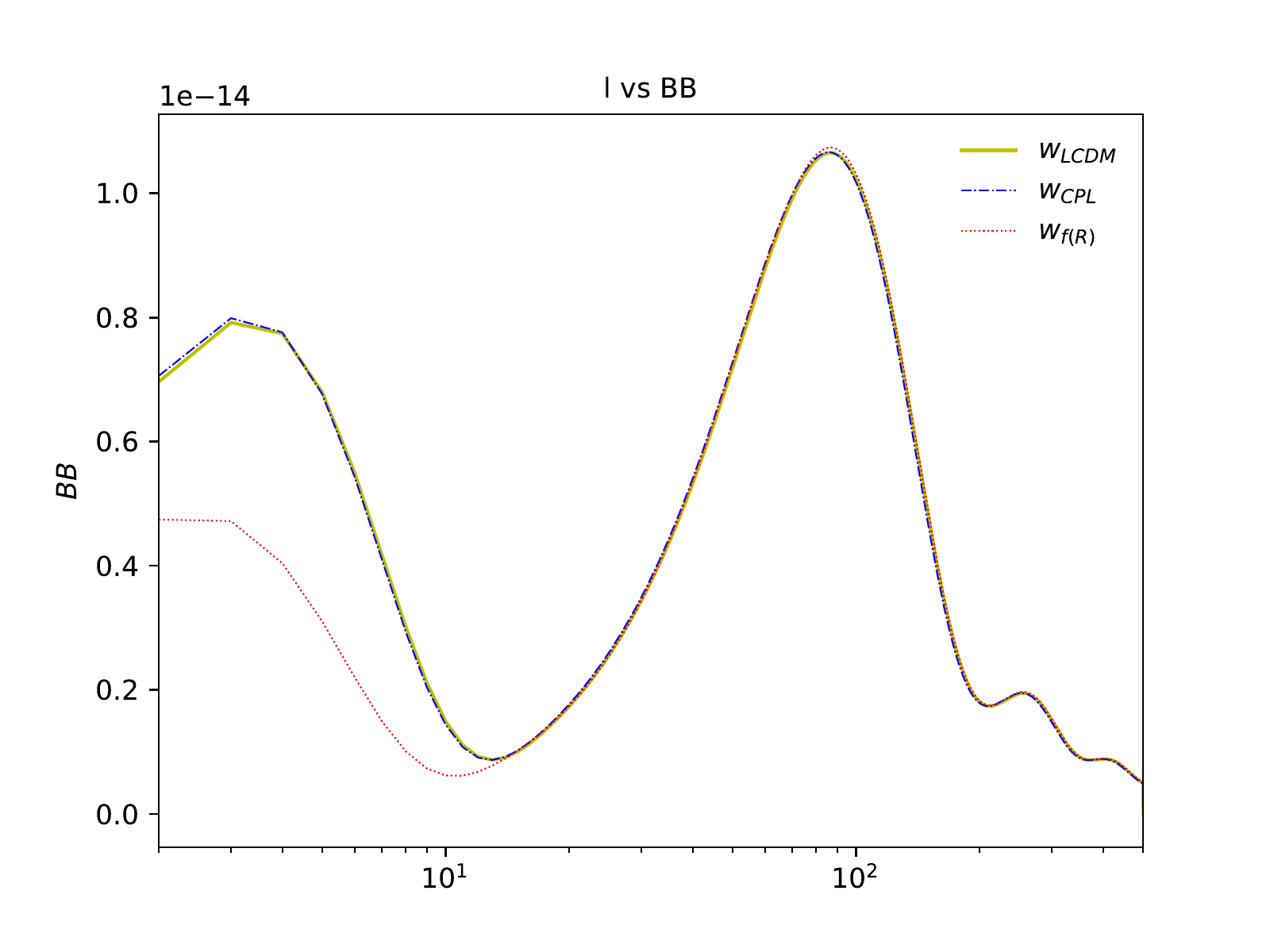}
	\includegraphics[width=0.45\textwidth]{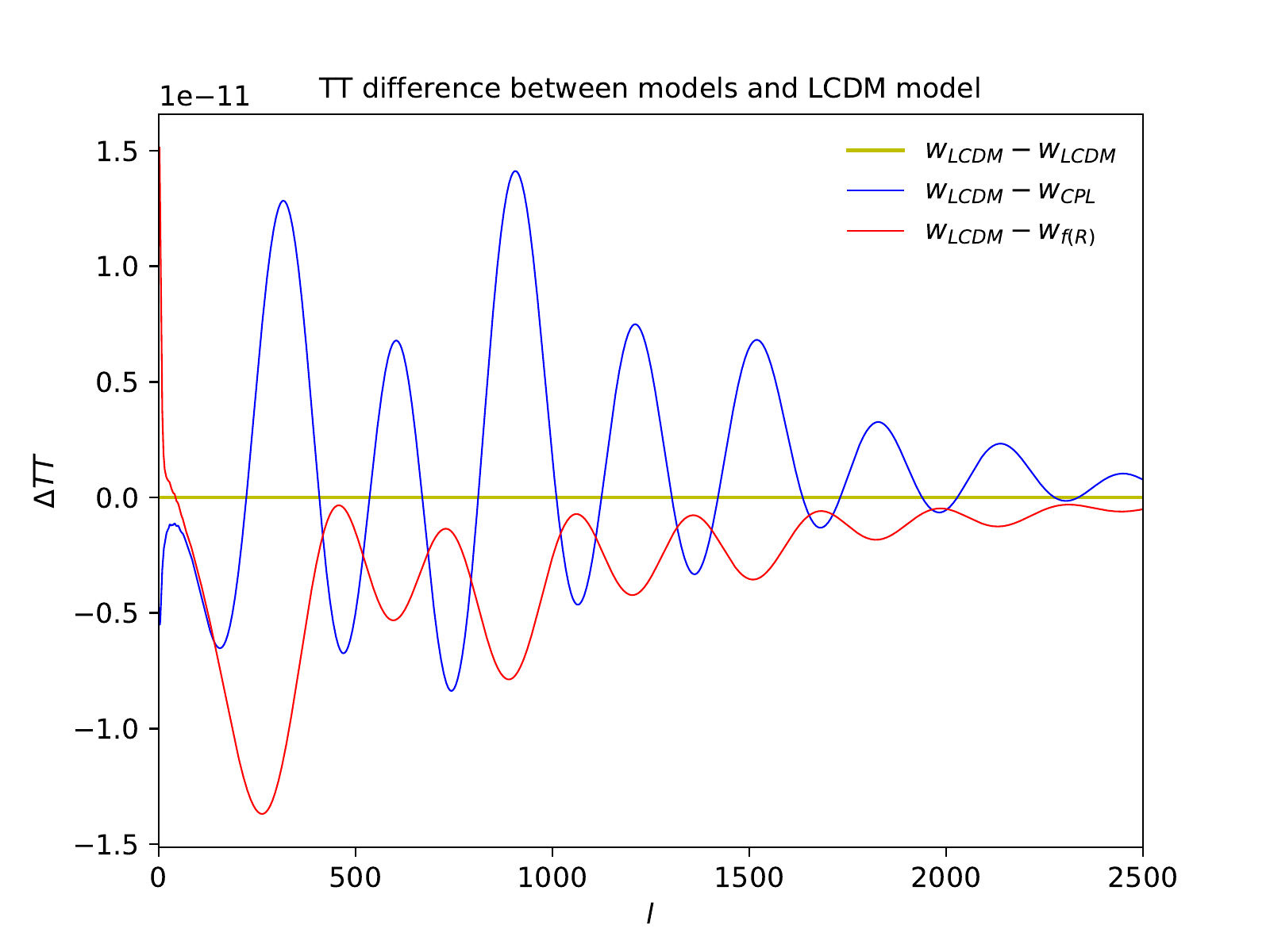}
	\includegraphics[width=0.45\textwidth]{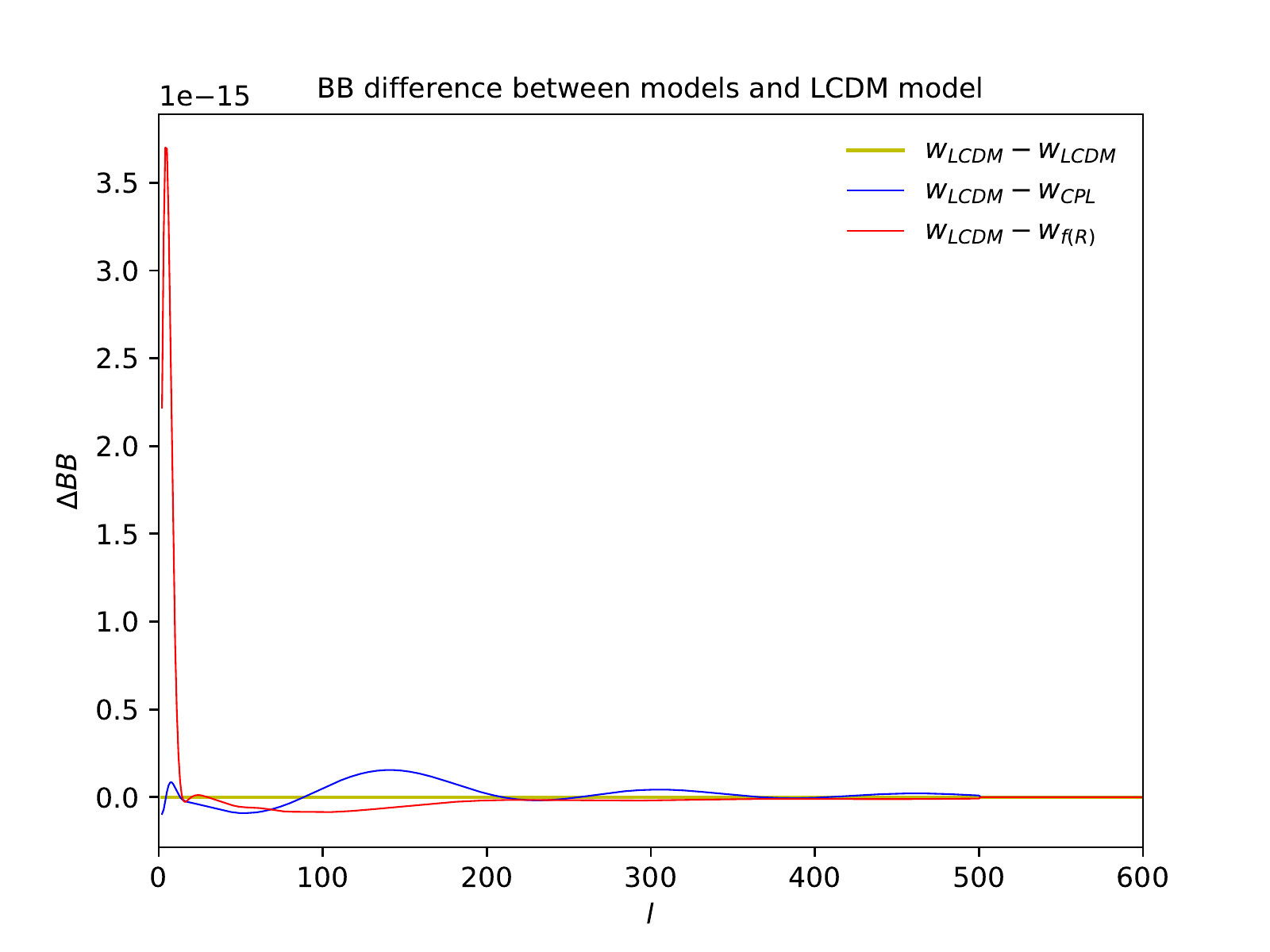}
	\caption{Power spectra for Exponential model derived from (\ref{eq:JJEeos}) and using the best-fits of the parameters obtained with the total sample (Pantheon SNeIa + BAO + CC + Planck 2018 + Ly-$\alpha$) reported in Table \ref{tab:total-exp}. We compare our parameterisation in this model with standard ones: $\Lambda$CDM and $CPL$ as $w(z) =w_0 +\frac{w_1 z}{ 1+z}$. \textit{Top (from Left to Right)}: TT and BB power spectrum. \textit{Bottom (from Left to Right)}: $\Delta$TT and $\Delta$BB ratio between standard models.}
	\label{fig:PS_Exponential}
\end{figure*}

For this model we obtain the C.L analysis in Figure \ref{fig:SpaceParameter_Exp_Et} with the bestfits reported in Table \ref{tab:exp-early} and their power spectra in Figure \ref{fig:PS_Exponential}

\begin{table*}
\centering
\caption{Best fits for the Exponential model using the early-time total sample (Planck 2018+Ly$\alpha$). From the final fit we obtain: $-\ln{\cal L}_\mathrm{min} =1407.05$ and $\chi^2=2814$.}
\begin{tabular}{|l|c|c|c|c|}
	\hline
	Parameter & Best-fit & mean$\pm\sigma$ & 95\% lower & 95\% upper \\ \hline
	$100~\omega_{b }$ &$2.254$ & $2.241_{-0.015}^{+0.015}$ & $2.238$ & $2.269$ \\
	$\omega_{cdm }$ &$0.1198$ & $0.1199_{-1.455e-03}^{+1.455e-03}$ & $0.118$ & $0.121$ \\
	$w_{0 }$ &$0.0159$ & $0.02409_{-0.012}^{+0.012}$ & $0.003$ & $0.028$ \\
	$H_{0 }$ &$68.2$ & $67.96_{-0.53}^{+0.79}$ & $67.67$ & $68.99$ \\
	$ln10^{10}A_{s }$ &$3.05$ & $3.041_{-0.016}^{+0.016}$ & $3.033$ & $3.066$ \\
	$n_{s }$ &$0.9653$ & $0.9644_{-0.0048}^{+0.0033}$ & $0.960$ & $0.968$ \\
	$\tau_{reio }$ &$0.05567$ & $0.05253_{-0.0077}^{+0.0072}$ & $0.047$ & $0.062$ \\
	$\Omega_{m }$ &$0.3074$ & $0.3081_{-0.043}^{+0.043}$ & $0.264$ & $0.350$ \\
	$\Omega0_{fld }$ &$0.6939$ & $0.6918_{-8.832}^{+8.832}$ & $0.685$ & $0.702$ \\
	$Y_{p }$ &$0.2479$ & $0.2479_{-6.467e-05}^{+6.467e-05}$ & $0.247$ & $0.248$ \\
	$\sigma8$ &$0.8265$ & $0.8217_{-7.720e-03}^{+7.720e-03}$ & $0.818$ & $0.834$ \\
	\hline
\end{tabular} \label{tab:exp-early}
 \end{table*}

\section{Results and Conclusions}
\label{sec:conclusions}

\begin{figure*}[ht]
	\centering
	\includegraphics[width=0.75\textwidth]{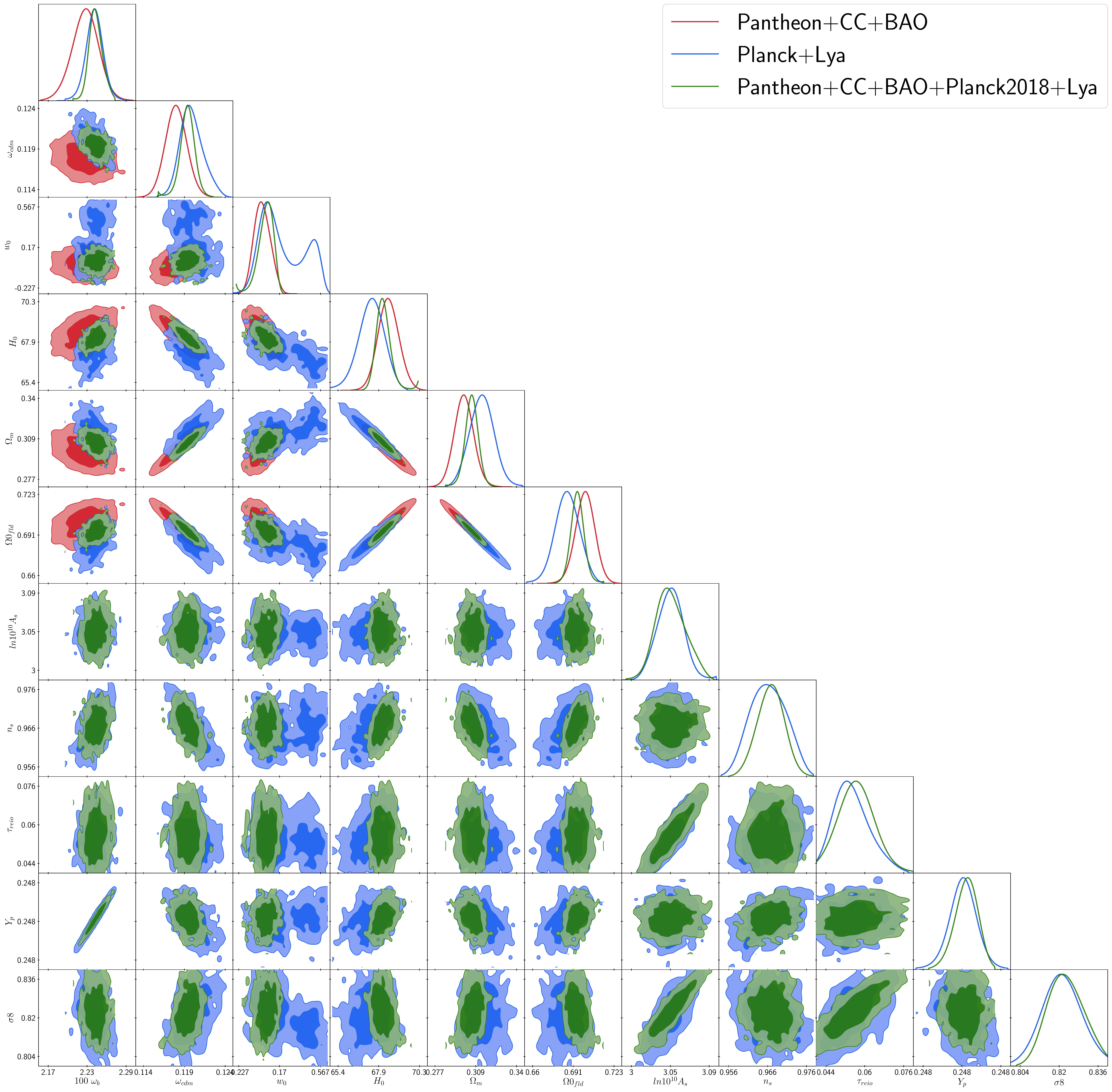}
	\caption{2-$\sigma$ C.L for the Hu-Sawicki model using the full total sample (SNeIa Pantheon + CC + BAO + Planck 2018+Ly$\alpha$) denoted by green color. The red color C.L corresponds to the sample Pantheon+CC+BAO and the blue color C.L corresponds to Planck 2018+Ly-$\alpha$.
}
	\label{fig:SpaceParameter_HS_3m}
\end{figure*}

In this work we have study a generic $f(R)$ parameterisation that can reproduce three cosmological viable models. The so-called adaptation of the JJE parameterisation (\ref{eq:fit}), iis an easy and efficient way to implement $f(R)$ gravity theories in any kind of survey and forecast, even at linear perturbative level. We perform general analysis by using the full sample (SNeIa Pantheon + CC + BAO + Planck 2018 +Ly$\alpha$). Our results are reported in Figures \ref{}-\ref{fig:SpaceParameter_HS_3m}-\ref{fig:SpaceParameter_Exp_3m} and Tables \ref{tab:total-Starobinsky}-\ref{tab:total-HuSawicki}-\ref{tab:total-exp}. As part of these analyses, we compute the tension between models using \cite{Camarena:2018nbr}
\begin{equation}
T_{H0}=\frac{|H_{0}-H_{0}^{\text{survey}}|}{\sqrt{\sigma^{2}_{H0}+\sigma^{2}_{\text{survey}}}},
\end{equation}
where \textit{surveys} are the $H_0$ prior from Planck 2018 and $\text{R}^{[18]}$used. Our results are reported in Tables \ref{tab:tensions}.  Using our $w_{f(R)}$ proposal the tension at this level seems to be alleviated for Hu-Sawicki model using early-time samples, while Exponential model relax the tension at late-times.
Notice that:
with the full total sample, the three $f(R)$ models in the $H_0$-$w_0$ phase space seems to have always a negative correlation. Using Planck 2018 + Ly-$\alpha$ sample ($H_0$-$w_0$ phase space): the Exponential model has a no convergence, while Starobinsky and Hu-Sawicki model has a negative correlation. With SN+CC+BAO sample ($H_0$-$w_0$ phase space), the Exponential model has a negative correlation and Starobinsky and Hu-Sawicki model has a negative one   

\begin{figure*}[ht]
	\centering
	\subfloat[]{
		\label{f:Yhe_st}
		\includegraphics[width=0.4\textwidth]{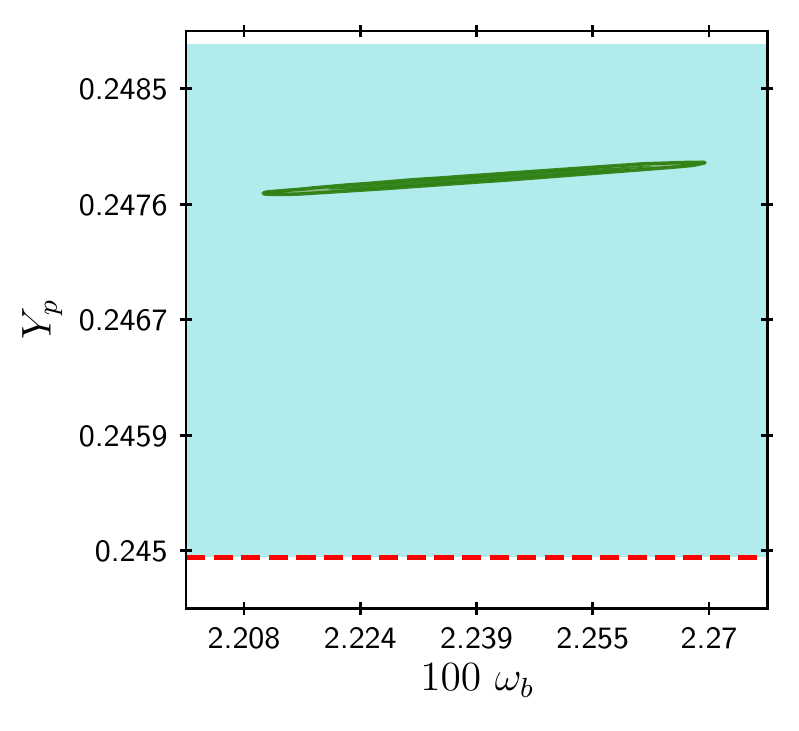}}
	\subfloat[]{
		\label{f:H0_st}
		\includegraphics[width=0.4\textwidth]{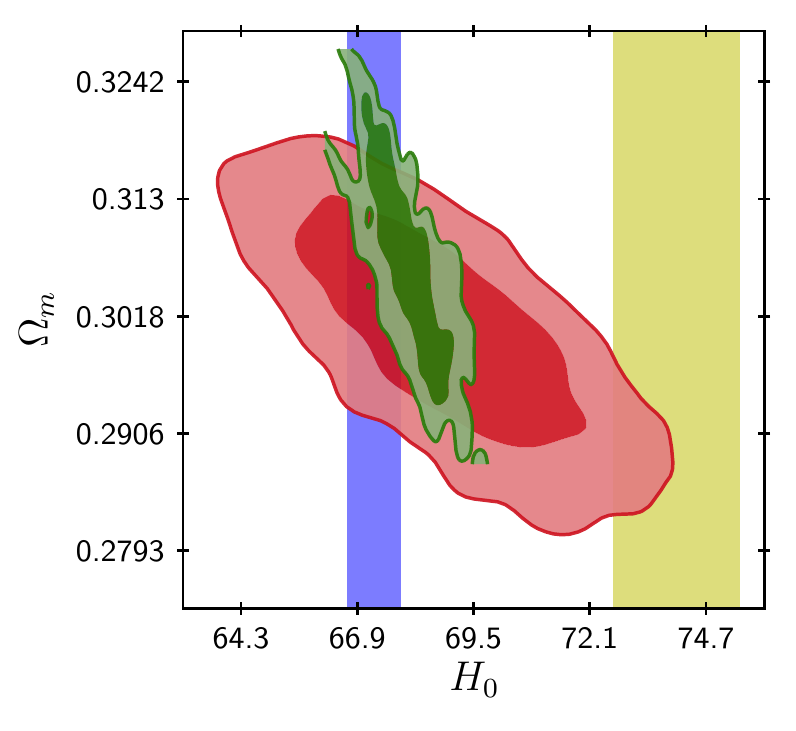}}
		\caption{C.L. Analysis for Starobinsky model with Planck 2018+Ly-$\alpha$. \textit{Left:} The red line denotes the C.L given by \cite{Aver:2015iza} $Y_{P}=0.2449$. The blue light region denotes the uncertainty top limit of  $\sigma_{Y_{P}}=+0.0040$. \textit{Right:} The red C.L denotes the model using late-time observations SNeIa Pantheon+CC+BAO and the green C.L denotes the model using early-time observations Planck 2018+Ly-$\alpha$. The purple region indicate the C.L using the Planck prior $H_0=67.27 \pm 0.60 \mathrm{km \ s^{-1} \ Mpc^{-1}}$ and  the yellow region indicates the C.L using the Cefeids $R^{[18]}$ prior $H_0=74.03 \pm 1.42 \mathrm{km \ s^{-1} \ Mpc^{-1}}$
}
	\label{fig:BBN_St}
\end{figure*}

\begin{figure*}[ht]
	\centering
	\subfloat[]{
		\label{f:Yhe_hs}
		\includegraphics[width=0.4\textwidth]{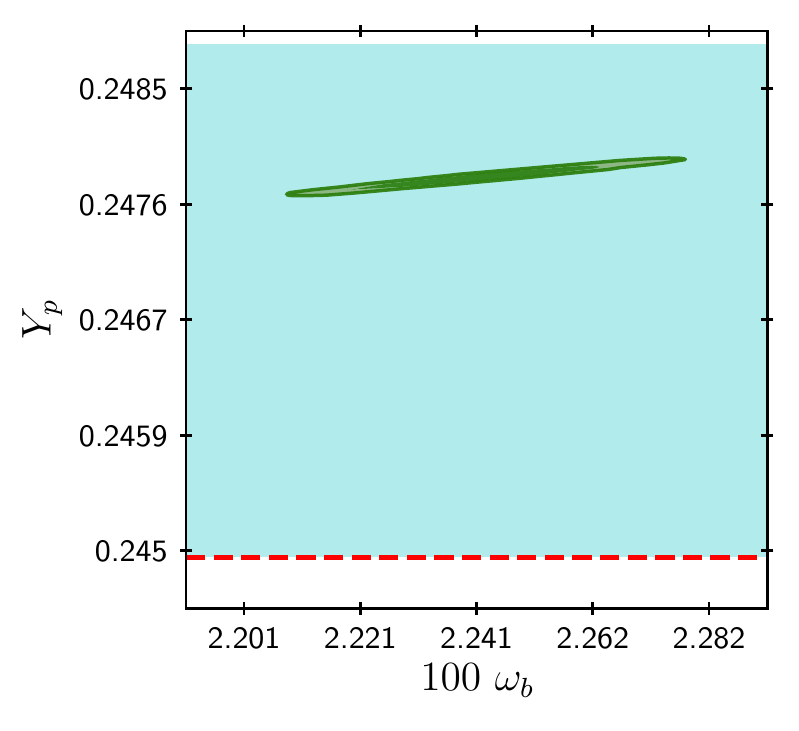}}
	\subfloat[]{
		\label{f:H0_hs}
		\includegraphics[width=0.4\textwidth]{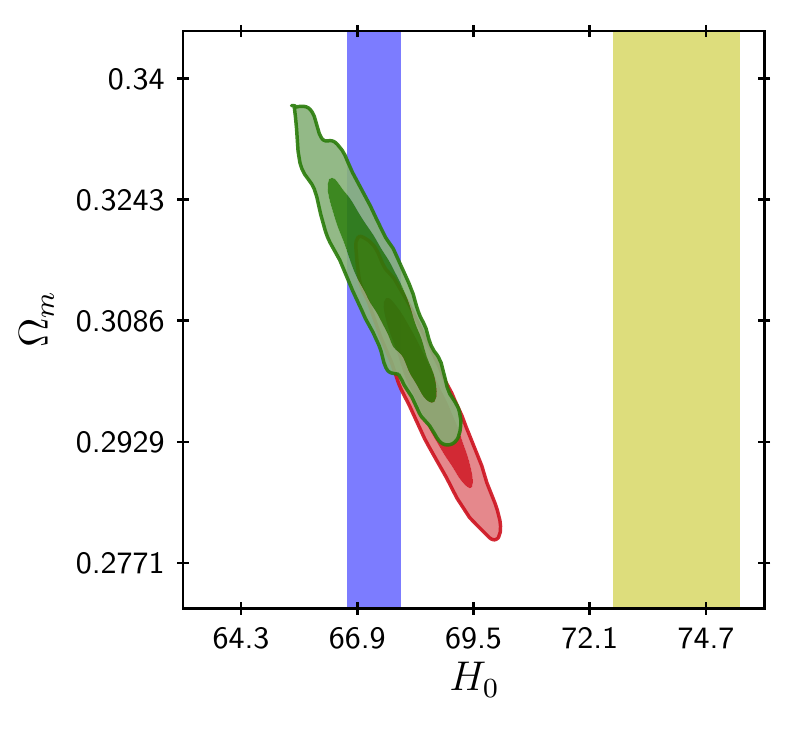}}
	\caption{
	C.L. Analysis for Hu-Sawicki model with Planck 2018+Ly-$\alpha$. \textit{Left:} The red line denotes the C.L given by \cite{Aver:2015iza} $Y_{P}=0.2449$. The blue light region denotes the uncertainty top limit of  $\sigma_{Y_{P}}=+0.0040$. \textit{Right:} The red C.L denotes the model using late-time observations SNeIa Pantheon+CC+BAO and the green C.L denotes the model using early-time observations Planck 2018+Ly-$\alpha$. The purple region indicate the C.L using the Planck prior $H_0=67.27 \pm 0.60 \mathrm{km \ s^{-1} \ Mpc^{-1}}$ and  the yellow region indicates the C.L using the Cefeids $R^{[18]}$ prior $H_0=74.03 \pm 1.42 \mathrm{km \ s^{-1} \ Mpc^{-1}}$}
	\label{fig:BBN_HS}
\end{figure*}

\begin{figure*}[ht]
	\centering
	\subfloat[]{
		\label{f:Yhe_exp}
		\includegraphics[width=0.4\textwidth]{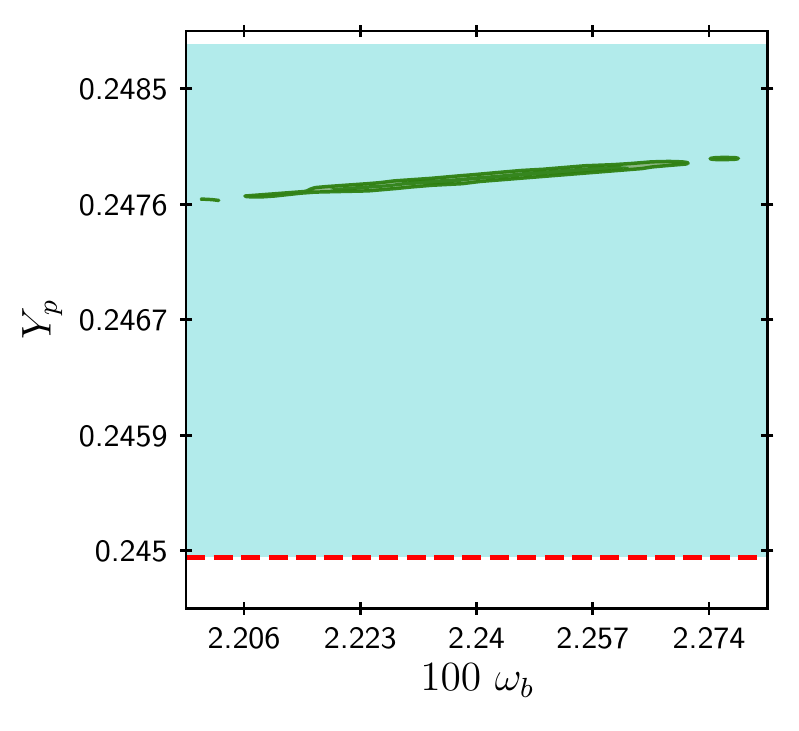}}
	\subfloat[]{
		\label{f:H0_exp}
		\includegraphics[width=0.4\textwidth]{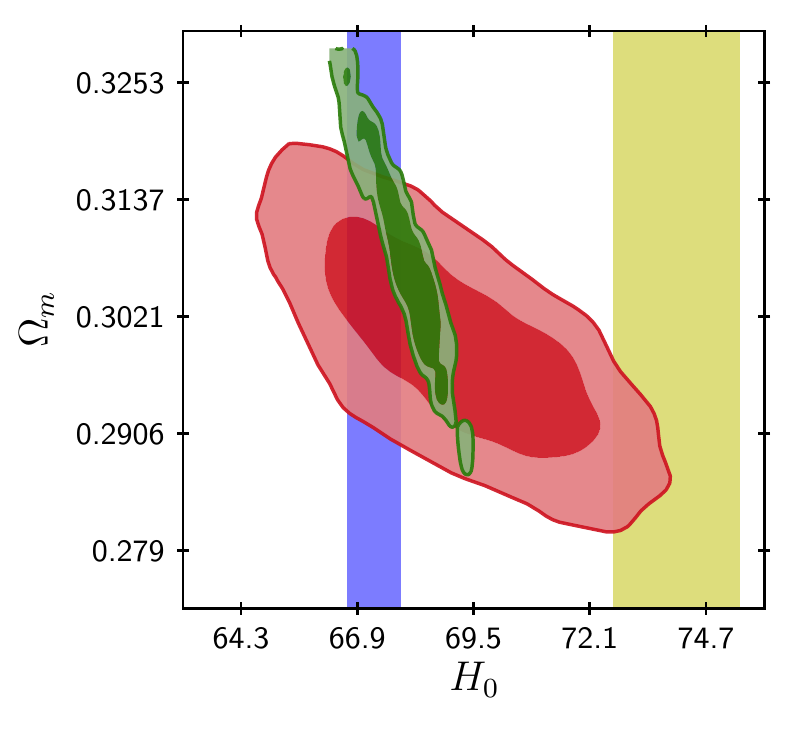}}
	\caption{C.L. Analysis for Exponential model with Planck 2018+Ly-$\alpha$. \textit{Left:} The red line denotes the C.L given by \cite{Aver:2015iza} $Y_{P}=0.2449$. The blue light region denotes the uncertainty top limit of  $\sigma_{Y_{P}}=+0.0040$. \textit{Right:} The red C.L denotes the model using late-time observations SNeIa Pantheon+CC+BAO and the green C.L denotes the model using early-time observations Planck 2018+Ly-$\alpha$. The purple region indicate the C.L using the Planck prior $H_0=67.27 \pm 0.60 \mathrm{km \ s^{-1} \ Mpc^{-1}}$ and  the yellow region indicates the C.L using the Cefeids $R^{[18]}$ prior $H_0=74.03 \pm 1.42 \mathrm{km \ s^{-1} \ Mpc^{-1}}$}
	\label{fig:BBN_Exp}
\end{figure*}

Another interesting feature is presented in Figures \ref{fig:BBN_St}-\ref{f:Yhe_hs}-\ref{f:Yhe_exp}. We found that our three $f(R)$ models lies in the C.L region of the primordial helium abundance, which is consistent with the BBN result $Y_p =0.2470\pm 0.0002$ from Planck 2018. Also we show that Starobinsky and Hu-Sawicki cosmologies using Planck 2018 +Ly$\alpha$ data are in agreement with the early-times constraints, while for a clear tension is given when late-time samplers are used. Furthermore, notice how the Exponential model lies near 2-$\sigma$ of C.L to $\text{R}^{[18]}$ current value.  With Hu-Sawicki cosmologies we can notice a higher deviation in the TT spectrum and in comparison to the other two models, which is evident from its $\Delta TT$. On the other hand, for the BB spectrum, Hu-Sawicki model at $10^{2}$ shows a deviation in comparison to $\Lambda$CDM and the other $f(R)$ models.

At linear perturbative level, a notorious difference in the $\Delta$TT spectra is present for our three $f(R)$ cosmologies in comparison to standard models as $\Lambda$CDM and CPL parameterisation. Moreover, using our $w_{f(R)}$ proposal the tension, at this level, seems to be alleviated as well.

%
%

\begin{figure*}[ht]
	\centering
	\includegraphics[width=0.75\textwidth]{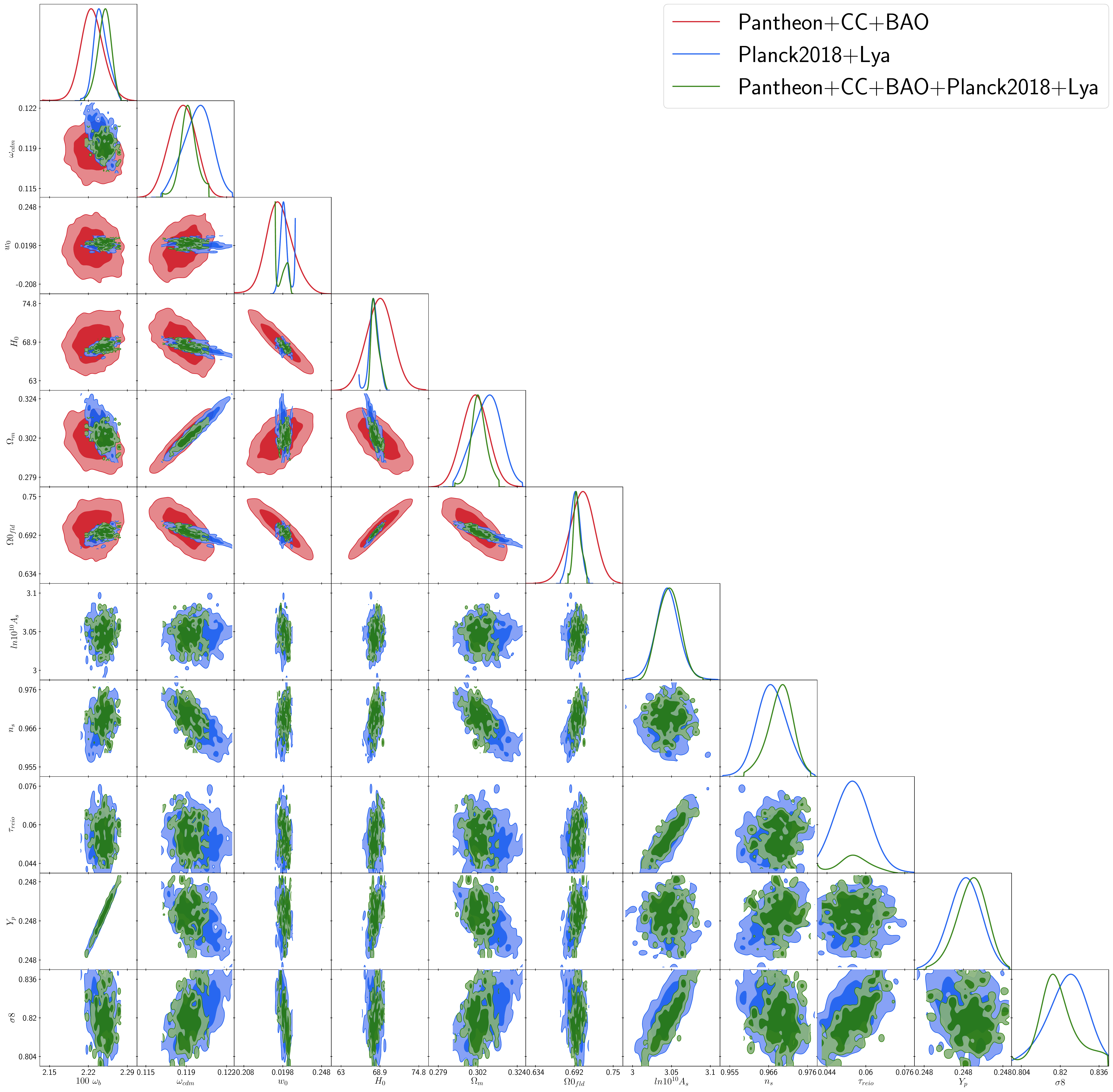}
	\caption{2-$\sigma$ C.L for the Starobinsky model using the full total sample (SNeIa Pantheon + CC + BAO + Planck 2018+Ly$\alpha$) denoted by green color. The red color C.L corresponds to the sample Pantheon+CC+BAO and the blue color C.L corresponds to Planck 2018+Ly-$\alpha$.
}
	\label{fig:SpaceParameter_Exp_3m}
\end{figure*}

\begin{figure*}[ht]
	\centering
	\includegraphics[width=0.75\textwidth]{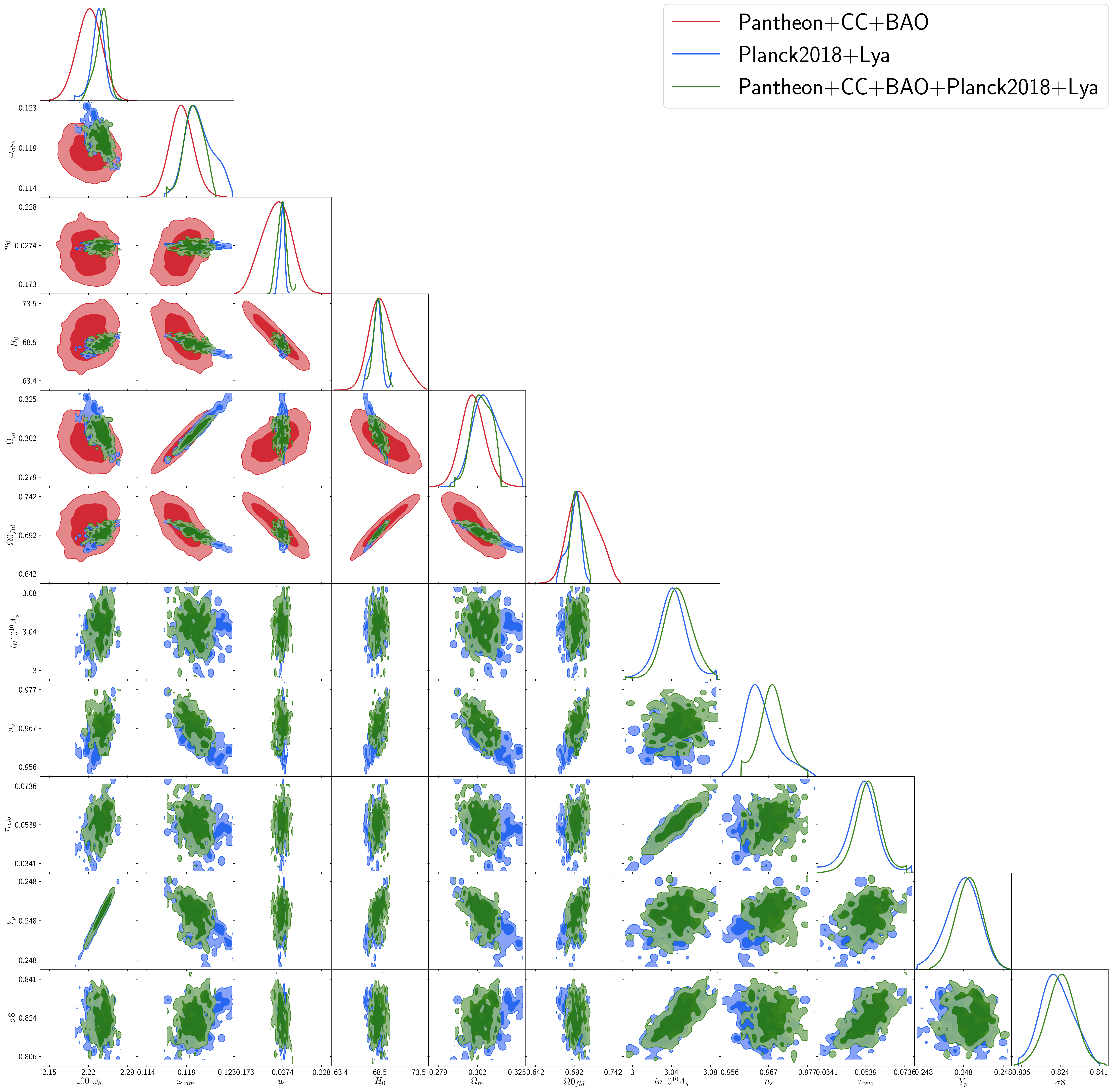}
	\caption{2-$\sigma$ C.L for the Exponential model using the full total sample (SNeIa Pantheon + CC + BAO + Planck 2018+Ly$\alpha$) denoted by green color. The red color C.L corresponds to the sample Pantheon+CC+BAO and the blue color C.L corresponds to Planck 2018+Ly-$\alpha$.
}
	\label{fig:SpaceParameter_Exp_3m}
\end{figure*}


\begin{table*}
\centering
\caption{Best fits for the Starobinsky model using the total sample (Pantheon SNeIa + BAO + CC + Planck 2018 + Ly-$\alpha$). From the final fit we obtain: $-\ln{\cal L}_\mathrm{min} =1929.38$ and $\chi^2=3859$.}
\begin{tabular}{|l|c|c|c|c|}
	\hline
	Parameter & Best-fit & mean$\pm\sigma$ & 95\% lower & 95\% upper \\ \hline
	$100~\omega_{b }$ &$2.263$ & $2.246_{-0.012}^{+0.012}$ & $2.250$ & $2.275$ \\
	$\omega_{cdm }$ &$0.1187$ & $0.1188_{-8.169e-04}^{+8.169e-04}$ & $0.117$ & $0.119$ \\
	$w_{0 }$ &$0.06792$ & $0.03244_{-0.020}^{+0.020}$ & $0.047$ & $0.088$ \\
	$H_{0 }$ &$67.72$ & $68.2_{-0.560}^{+0.560}$ & $67.159$ & $68.280$ \\
	$ln10^{10}A_{s }$ &$3.045$ & $3.044_{-0.014}^{+0.014}$ & $3.030$ & $3.059$ \\
	$n_{s }$ &$0.9698$ & $0.9687_{-3.607e-03}^{+3.607e-03}$ & $0.966$ & $0.973$ \\
	$\tau_{reio }$ &$0.05599$ & $0.0552_{-6.533e-03}^{+6.533e-03}$ & $0.049$ & $0.062$ \\
	$\Omega_{m }$ &$0.3006$ & $0.3023_{-4.566e-03}^{+4.566e-03}$ & $0.296$ & $0.305$ \\
	$\Omega0_{fld }$ &$0.6917$ & $0.6961_{-6.052e-03}^{+6.052e-03}$ & $0.685$ & $0.697$ \\
	$Y_{p }$ &$0.2479$ & $0.2479_{-5.432e-05}^{+5.432e-05}$ & $0.247$ & $0.248$ \\
	$\sigma8$ &$0.8134$ & $0.8192_{-8.310e-03}^{+8.310e-03}$ & $0.805$ & $0.821$ \\
	\hline
\end{tabular} \label{tab:total-Starobinsky}
 \end{table*}

\begin{table*}
\centering
\caption{Best fits for the Hu-Sawicki model using the total sample (Pantheon SNeIa + BAO + CC + Planck 2018 + Ly-$\alpha$). From the final fit we obtain: $-\ln{\cal L}_\mathrm{min} =1930.4$ and $\chi^2=3861$.}
\begin{tabular}{|l|c|c|c|c|}
\hline
	Parameter & Best-fit & mean$\pm\sigma$ & 95\% lower & 95\% upper \\ \hline
	$100~\omega_{b }$ &$2.254$ & $2.244_{-0.012}^{+0.012}$ & $2.242$ & $2.266$ \\
	$\omega_{cdm }$ &$0.1191$ & $0.1194_{-0.00082}^{+0.00097}$ & $0.1176$ & $0.1212$ \\
	$w_{0 }$ &$0.1006$ & $0.03188_{-0.06}^{+0.075}$ & $0.040$ & $0.1756$ \\
	$H_{0 }$ &$68.16$ & $68.15_{-0.49}^{+0.4}$ & $67.31$ & $69$ \\
	$ln10^{10}A_{s }$ &$3.046$ & $3.047_{-0.019}^{+0.014}$ & $3.013$ & $3.082$ \\
	$n_{s }$ &$0.9729$ & $0.9668_{-0.0037}^{+0.0036}$ & $0.9598$ & $0.9737$ \\
	$\tau_{reio }$ &$0.0541$ & $0.05588_{-0.0087}^{+0.0074}$ & $0.03973$ & $0.07413$ \\
	$\Omega_{m }$ &$0.3048$ & $0.3054_{-0.0054}^{+0.006}$ & $0.2944$ & $0.3164$ \\
	$\Omega0_{fld }$ &$0.6951$ & $0.6945_{-0.006}^{+0.0054}$ & $0.6835$ & $0.7055$ \\
	$Y_{p }$ &$0.2479$ & $0.2479_{-5.1e-05}^{+5.1e-05}$ & $0.247$ & $0.248$ \\
	$\sigma8$ &$0.8224$ & $0.823_{-0.0083}^{+0.007}$ & $0.8078$ & $0.8384$ \\
	\hline
\end{tabular} \label{tab:total-HuSawicki}
 \end{table*}

\begin{table*}
\centering
\caption{Best fits for the Exponential model using the total sample (Pantheon SNeIa + BAO + CC + Planck 2018 + Ly-$\alpha$). From the final fit we obtain: $-\ln{\cal L}_\mathrm{min} =1928.56$ and $\chi^2=3857$.}

\begin{tabular}{|l|c|c|c|c|}
	\hline
	Parameter & Best-fit & mean$\pm\sigma$ & 95\% lower & 95\% upper \\ \hline
	$100~\omega_{b }$ &$2.239$ & $2.248_{-0.011}^{+0.014}$ & $2.222$ & $2.269$ \\
	$\omega_{cdm }$ &$0.1195$ & $0.1193_{-1.015e-03}^{+1.015e-03}$ & $0.118$ & $0.120$ \\
	$w_{0 }$ &$0.06334$ & $0.02126_{-0.022}^{+0.022}$ & $0.041$ & $0.085$ \\
	$H_{0 }$ &$67.14$ & $68.3_{-0.63}^{+0.66}$ & $66.51$ & $67.8$ \\
	$ln10^{10}A_{s }$ &$3.06$ & $3.046_{-0.015}^{+0.015}$ & $3.044$ & $3.075$ \\
	$n_{s }$ &$0.9665$ & $0.9677_{-3.646e-03}^{+3.646e-03}$ & $0.962$ & $0.970$ \\
	$\tau_{reio }$ &$0.06303$ & $0.05541_{-7.423e-03}^{+7.423e-03}$ & $0.055$ & $0.070$ \\
	$\Omega_{m }$ &$0.308$ & $0.3045_{-5.884e-03}^{+-5.884e-03}$ & $0.302$ & $0.313$ \\
	$\Omega0_{fld }$ &$0.6851$ & $0.6959_{-7.008e-03}^{+7.008e-03}$ & $0.678$ & $0.692$ \\
	$Y_{p }$ &$0.2478$ & $0.2479_{-4.6e-05}^{+5.7e-05}$ & $0.2478$ & $0.248$ \\
	$\sigma8$ &$0.8227$ & $0.8237_{-7.247e-03}^{+7.247e-03}$ & $0.815$ & $0.829$ \\
	\hline
\end{tabular}  \label{tab:total-exp}
 \end{table*}


\begin{table*} \label{tab:tensions}
\begin{center}
\caption{Tension values for $w_{f(R)}$ parameterisation: \textit{Top:} Starobinsky model. \textit{Middle:} Hu-Sawicki model. \textit{Bottom:} Exponential model. \textit{First column:} Riess et al ($\text{R}^{[18]}$) versus  Planck+Ly$\alpha$. \textit{Second column:} Riess et al ($\text{R}^{[18]}$) versus vs SNeIa(Pantheon)+CC+BAO. \textit{Third column:} SNeIa(Pantheon)+CC+BAO versus Planck+Ly$\alpha$.
}
\begin{tabular}{|c|c|c|c|c|c|c|c|c|} 
 \hline
\multicolumn{3}{ |c| }{\textbf{$\text{R}^{[18]}$ vs Planck+Ly$\alpha$}} & \multicolumn{3}{ |c| }{\textbf{$\text{R}^{[18]}$ vs SNeIa+CC+BAO}} & \multicolumn{3}{ |c| }{\textbf{SNeIa+CC+BAO vs Planck+Ly$\alpha$}} \\ \hline
$H_{0}$ & $H_0$ & $T_{H0}$ & $H_0$ & $H_0$ & $T_{H0}$ & $H_0$ & $H_0$ & $T_{H0}$ \\\hline
$74.03$ & $67.72$ & $4.0456$ & $74.03$ & $68.75$ & $2.0828$ & $68.75$ & $67.72$ & $0.4688$ \\
\hline
 \end{tabular} \label{tab:tensions}
 \end{center}
 
\begin{center}
\begin{tabular}{|c|c|c|c|c|c|c|c|c|} 
 \hline
\multicolumn{3}{ |c| }{\textbf{$\text{R}^{[18]}$ vs Planck+Ly$\alpha$}} & \multicolumn{3}{ |c| }{\textbf{$\text{R}^{[18]}$ vs SNeIa+CC+BAO}} & \multicolumn{3}{ |c| }{\textbf{SNeIa+CC+BAO vs Planck+Ly$\alpha$}} \\ \hline
$H_{0}$ & $H_0$ & $T_{H0}$ & $H_0$ & $H_0$ & $T_{H0}$ & $H_0$ & $H_0$ & $T_{H0}$ \\\hline
$74.03$ & $67.6$ & $3.46$ & $74.03$ & $68.41$ & $3.5597$ & $68.41$ & $67.6$ & $0.513$ \\
\hline
 \end{tabular} 
 \end{center}
 
\begin{center}
\begin{tabular}{|c|c|c|c|c|c|c|c|c|} 
 \hline
\multicolumn{3}{ |c| }{\textbf{$\text{R}^{[18]}$ vs Planck+Ly$\alpha$}} & \multicolumn{3}{ |c| }{\textbf{$\text{R}^{[18]}$ vs SNeIa+CC+BAO}} & \multicolumn{3}{ |c| }{\textbf{SNeIa+CC+BAO vs Planck+Ly$\alpha$}} \\ \hline
$H_{0}$ & $H_0$ & $T_{H0}$ & $H_0$ & $H_0$ & $T_{H0}$ & $H_0$ & $H_0$ & $T_{H0}$ \\\hline
$74.03$ & $68.2$ & $3.587$ & $74.03$ & $68.86$ & $1.9744$ & $68.86$ & $68.2$ & $0.282$ \\
\hline
 \end{tabular} 
 \end{center}
 \end{table*}


\section*{Acknowledgements}
CE-R acknowledges the Royal Astronomical Society as FRAS 10147, the support 
by PAPIIT Project IA100220 and acknowledge networking support by the COST Action CA18108.
N. Jim\'enez thank Consejo Nacional de
Ciencia y Tecnolog\'ia (CONACyT) for her master fellowship.


\end{document}